\documentclass[submission, Phys]{SciPost}

\hypersetup{
	colorlinks,
	linkcolor={red!50!black},
	citecolor={blue!50!black},
	urlcolor={blue!80!black}
}

\usepackage[bitstream-charter]{mathdesign}
\DeclareSymbolFont{usualmathcal}{OMS}{cmsy}{m}{n}
\DeclareSymbolFontAlphabet{\mathcal}{usualmathcal}

\begin{document}

%\title[Molecular spinning tops]{Cooperative dynamics in two-component out-of-equilibrium systems: Molecular "spinning tops" 
%}

\begin{center} \Large \textbf{ Cooperative dynamics in two-component out-of-equilibrium systems: Molecular "spinning tops" }
\end{center}
\begin{center} 
Victor S Dotsenko$^1$, Pascal Viot$^1$, Alberto Imparato$^2$ \& Gleb Oshanin$^1$
\end{center}

\begin{center}
$^1$Sorbonne Universit\'e, CNRS, Laboratoire de Physique Th\'eorique de la Mati\`ere Condens\'ee (UMR CNRS 7600), 4 Place Jussieu, 75252 Paris Cedex 05, France\\
$^2$Department of Physics and Astronomy, Aarhus University, 
		Ny Munkegade 120, 
		8000 Aarhus C,
		Denmark
\end{center}
	
%\author{Viktor S Dotsenko$^1$, Pascal Viot$^1$, Alberto Imparato$^2$ \& Gleb Oshanin$^1$}

\begin{center}
	\today
\end{center}

%\address{$^1$Sorbonne Universit\'e, CNRS, Laboratoire de Physique Th\'eorique de la Mati\`ere Condens\'ee (UMR CNRS 7600), 4 Place Jussieu, 75252 Paris Cedex 05, France}
%\address{$^2$Department of Physics and Astronomy, Aarhus University, 
%Ny Munkegade 120, 
%8000 Aarhus C,
%Denmark}

\section*{Abstract}
%\begin{abstract}
We study the two-dimensional Langevin dynamics of a 
two-component system,  
whose components are in contact with heat baths kept at different temperatures.
Dynamics is constrained by an optical trap
and   
 the \text{dissimilar} species interact  via a quadratic potential. 
We realize that the system evolves towards a peculiar
non-equilibrium steady-state with a non-zero probability current possessing a non-zero curl, 
such that the randomly moving particles are spinning around themselves, like "spinning top" toys.  
Our analysis shows that the spinning motion is correlated and also
reveals an emerging cooperative behavior of the spatial components of the probability currents of dissimilar species.
%{\bf We study the Langevin dynamics of particles of two kinds present at equal amounts $N$,
%that are all held by an optical tweezer, i.e., each particle is "connected" 
%to the origin by a harmonic spring of strength $\kappa$. Each kind of particles lives at its own temperature
%and the \text{dissimilar} species interact between themselves via a quadratic potential $U(r) = - u_0 + u r^2/2$, where  $u_0$ is  a constant,  
%$r$ is the inter-particle distance in a pair of particles, and $u$ is the amplitude of interactions, which can be repulsive ($u<0$) or attractive ($u>0$). 
%We show analytically that for $-\kappa/(2N) < u < \infty$, (excluding $u=0$), such a system evolves towards a non-equilibrium steady-state with a non-vanishing probability current. The latter possesses a non-zero curl, implying that the particles which move randomly within the optical trap are also spinning around themselves, like "spinning top" toys.  Strength and direction of spinning depends on particles' positions and is synchronized. Moreover, our analysis reveals an emerging cooperative behaviour of the components of the probability currents of dissimilar species, which resembles the one observed in the Brownian Gyrator model.}
%\end{abstract}

Keywords: Out-of-equilibrium dynamics, random rotational motion, molecular motors

\vspace{10pt}
\noindent\rule{\textwidth}{1pt}
\tableofcontents\thispagestyle{fancy}
\noindent\rule{\textwidth}{1pt}
\vspace{10pt}

%\maketitle

%%%%%%%%%%%%%%%%%%%%%%%%%%%%%%%%%%%%%%%%%%%%%%%%%%%%%%%%%%%%%%%%
\section{Introduction}

A directional (translational or rotational) motion of small particles or molecules  emerging due 
to a breaking of some symmetries and the ensuing effective rectification of random fluctuations, 
has been in focus of research for many years.  A variety of analytical, numerical and experimental results has been obtained, as well as some general  
concepts and approaches have been developed in a series of seminal papers (see, e.g., \cite{prost,faucheaux,duke,strick,sawtooth1,sawtooth2,sawtooth3,sawtooth4}). Advances made in this field were reviewed in \cite{julicher,reimann,hanggi,astumian}.

Here we are specifically concerned with spontaneous \textit{rotational} motion which emerges due to broken symmetries. The molecular motors which exhibit such a behavior are encountered in nature, e.g.,  appearing as an essential unit in pumping mechanisms of protons across the vacuolar membranes \cite{tsunoda}, or prompting a rotary motion of bacterial flagella \cite{berg}.
In some instances, such motors can be artificially designed and used 
as efficient pumps acting in the bulk and at the surfaces 
of liquids \cite{wang}, or micro windmills driven by tumbling bacteria \cite{angelani1,angelani2}.

A very simple but yet a non-trivial model that exhibits an erratic rotational motion  
is that of a Brownian Gyrator (BG) \cite{peliti,rei}.  It consists of two linearly-coupled Ornstein-Uhlenbeck processes
each subject to its own independent noise, such that it can be envisaged as a model for dynamics of a particle on a two-dimensional plane in presence of a parabolic potential and non necessarily equal temperatures
along the two cartesian coordinates. In case of a broken symmetry, i.e., when the temperatures are different, 
the system evolves towards a non-equilibrium steady-state with a non vanishing probability current. The latter possesses a non-zero curl centered at the origin, meaning that the
particle undergoes, on average, rotations around the origin. In case of equal temperatures the system evolves towards a thermal equilibrium state and the rotational motion is absent. Within the recent years, very diverse features and a behavior of pertinent properties
characterizing the BG have been discussed in the literature for delta-correlated noises \cite{cil1,cil2,crisanti,2,3,alberto,bae,13,Lam31,Lam32,cherail,sara,vulp,vik,olga,olga2} and even for the noises with long-range temporal correlations \cite{nasc,alessio}. 

In this paper, motivated in part by the analysis of a collective behavior of rotating motors on surfaces of biological membranes \cite{lenz} and especially by a notable more general theoretical analysis of the behavior of 
particles of different types that are in contact with thermostats kept at different temperatures \cite{gros}, we 
propose a novel minimalistic model which 
exhibits a non-trivial stochastic rotational motion whose characteristic features 
are in some aspects completely different  and in some other resemble the ones encountered 
 in  the BG model. 
 %Differently from the BG considered in, e.g., \cite{rei}, our model, in its most general form,  consists of $2N$ interacting particles, and the collective effects between them  play a fundamental role in the syncronized rotation that we present here.  
%{\alb analysis of the performance in the limits $N\to 1$ and $N\to \infty$ ?}
More specifically, we consider the two-dimensional Langevin dynamics of two types of point-like particles  - type $1$ and type $2$, which
   live at their own temperatures:  $T_1$ or $T_2$, respectively. The ensemble of particles
  is held by an optical tweezer;  that being, all the particles are tethered to the origin by harmonic springs with the same strength $\kappa$. Moreover, the dissimilar species interact between themselves via a quadratic potential of the form
  \begin{align}
  \label{potent}
  U(r) = - u_0 + \frac{u}{2} r^2,
  \end{align}
  where $r$ is the inter-particle distance in each pair of dissimilar particles, $u_0$ defines the depth of the potential well at the closest approach distance,   and $u$ is the amplitude which  can be positive (attractive interactions) or negative (repulsive interactions).  We proceed to show that for some values of $u$  such a system evolves towards a non-equilibrium steady-state with a non-vanishing probability current that possesses a non-zero curl, implying that the particles which move randomly within the optical trap are also spinning around themselves, like "spinning top" toys,  in contrast to the BG case in which rotation takes place around the origin.  Spinning motion of  particles is position-dependent and is synchronized, if we consider the dynamics of the centers of mass of the two ensembles.  The term "synchronized" signifies here that the value and the sign of the curl of the probability current associated with a single particle of type $1$ appearing at some fixed position (or the center of mass for an ensemble of particles $1$) 
  are the \textit{same} as for the corresponding property associated with the particle (or ensemble of particles) of type $2$ appearing at some fixed 
  position.
  
  We also show that in such a system the spatial components of the probability currents of the dissimilar species appear to exhibit a peculiar cooperative behavior: Namely, the stream plots of the components of currents of dissimilar particles reveal that the latter evolve along closed elliptic orbits centered around the origin,  
  which resembles in a way a behavior observed in the BG model, where the $x$- and $y$-components of the probability current of the BG form precisely such  a kind of patterns \cite{Lam31,Lam32}. As mentioned above, for the BG this signifies that the particle circulates spontaneously around the origin. In our setting,  such a rotation around the origin does not take place because it concerns the \textit{dissimilar} particles and therefore only points on the emerging correlations and the ensuing cooperativity.  Most of our analysis is performed for the simplest setting with just two particles. For the  case with $N$ particles we present the explicit formulas for the currents and their curls,  showing that they are not equal to zero in general, 
  and also discuss some peculiarities of the behavior in the limit $N \to \infty$.
  
Lastly, we remark that for the potential in eq. \eqref{potent} the model is exactly solvable. On intuitive grounds, one may also expect that our results will be still valid in the low-temperature regime for more general potentials which admit the form in eq. \eqref{potent} for small values of $r$: for small $T_1$ and $T_2$ both types of particles will reside close to the bottom of the potential well such that effectively all inter-particles distances will be small such that only the short-$r$ behavior of $U(r)$ will be important. Spinning of particles in other experimentally-relevant geometrical settings, in which each type of particles is held by their own tweezer focalized on some point in space and the coupling between dissimilar species is provided by the hydrodynamic interactions (see \cite{alb1}) will be studied elsewhere \cite{alb2}.
 %Most of our analysis pertains to the situation with only two particles but we also provide a generalization over the case %when two kinds of particles are present in some amounts.

The paper is structured as follows:  In Sec. \ref{sec:model} we define our model and introduce basic notations. 
In Sec. \ref{sec:2particles} we consider the simplest model with two particles and  
present our main analytical results.  
In Sec. \ref{sec:Npart} we consider the general case with two types of particles which are present at equal amounts $N$. We obtain the stationary probability distribution in this general case from which we derive explicit expressions for the particles currents and calculate their curls. 
%The model with 4 particles is studied in details 
%In Sec.\ref{sec:beyond}, in order to test the robustness of the phenomenon, we perform simulations for 
%interaction potential which deviates from the harmonic approximation, for which exact results have been obtained.
We conclude with a brief recapitulation of our results and a discussion in Sec. \ref{conc}. Details of intermediate calculations are relegated to \ref{sec:appA}.

\section{Model and basic notations}
\label{sec:model}

Consider a two-dimensional plane with just two particles -- particle $1$ and particle $2$, whose instantaneous positions are denoted by vectors
${\bf r}_1=(x_1, y_1)$ and ${\bf r}_2 = (x_2, y_2)$, respectively. The particles are tethered to the origin by a harmonic spring due to an optical trap, characterized by the stiffness constant $\kappa$, and experience attractive interaction between themselves 
via potential $U(r)$ in eq. \eqref{potent}, where $r = |{\bf r}_1 - {\bf r}_2|$ is the instantaneous distance between  the  
particles $1$ and $2$. 
We stipulate that 
the particles' dynamics obeys the coupled (via $U(r)$) Langevin equations of the form
\begin{equation}
\begin{split}
\label{langevin}
\nu \, \dot{\bf r}_1 &= - \kappa \, {\bf r}_1 -  \nabla_1 U(r)  + {\bf \zeta}_1 \,, \\
\nu \, \dot{\bf r}_2 &= - \kappa \,  {\bf r}_2 -  \nabla_2 U(r) + {\bf \zeta}_2 \,,
\end{split}
\end{equation}
where $\nu$ is the viscosity, the dot denotes the time derivative,  $ \nabla_1$ and $ \nabla_2$ stand for 
 the gradient operators with respect to the positions of  the particles, 
while ${\bf \zeta}_1$ and ${\bf \zeta}_2$ are mutually \textit{independent} zero-mean Gaussian noises with the covariance functions
\begin{equation}
\begin{split}
\label{noises}
\overline{{\bf \zeta}_1(t) {\bf \zeta}_1(t')} &=  2 \nu T_1 \delta(t-t') \,, \\
\overline{{\bf \zeta}_2(t) {\bf \zeta}_2(t')} &=  2 \nu T_2 \delta(t-t') \,. 
\end{split}
\end{equation}
In eqs. \eqref{noises},  the overbar denotes the average over thermal histories. 
 To ease the notations  and to avoid having too many parameters in what follows, we set $\nu$ equal to unity, which means that all the variables and parameters are appropriately rescaled. We do not introduce new notations for the rescaled parameters but simply suppose that they are all dimensionless.
 
 Further on,  for the Langevin eqs. \eqref{langevin} one can write down the associated Fokker-Planck equation obeyed by the probability density function $P_t({\bf r_1}|{\bf r_2})$. In the steady-state, the Fokker-Planck equation for $P({\bf r_1}|{\bf r_2}) = P_{t = \infty}({\bf r_1}|{\bf r_2})$ reads
 \begin{align}
 \label{fp}
{\rm div}_1 {\bf j}_1({\bf r_1}|{\bf r_2}) + {\rm div}_2 {\bf j}_2({\bf r_1}|{\bf r_2}) = 0 \,,
 \end{align}
 in which equation ${\rm div}$ denotes the divergence operator, while
  ${\bf j}_1({\bf r_1}|{\bf r_2}) = (j_1^x({\bf r_1}|{\bf r_2}), j_1^y({\bf r_1}|{\bf r_2}))$ and ${\bf j}_2({\bf r_1}|{\bf r_2}) =  (j_2^x({\bf r_1}|{\bf r_2}), j_2^y({\bf r_1}|{\bf r_2}))$ are the probability currents :
 \begin{equation}
 \begin{split}
 \label{j}
 {\bf j}_1({\bf r_1}|{\bf r_2}) &=  - T_1 \nabla_1 P({\bf r_1}|{\bf r_2}) - P({\bf r_1}|{\bf r_2}) \nabla_1 H({\bf r}_1|{\bf r}_2) \,, \\
 {\bf j}_2({\bf r_1}|{\bf r_2})  &=  - T_2 \nabla_2 P({\bf r_1}|{\bf r_2}) - P({\bf r_1}|{\bf r_2}) \nabla_2 H({\bf r}_1|{\bf r}_2) \,,
 \end{split}
 \end{equation}
 where 
for the potential in eq. \eqref{potent} the Hamiltonian $H({\bf r}_1|{\bf r}_2)$ is given explicitly by
\begin{align}
\label{H}
 H({\bf r}_1|{\bf r}_2) = \left(\kappa + u\right) \left(\frac{{\bf r}_1^2}{2} + \frac{{\bf r}_2^2}{2} - \lambda \, ({\bf r}_1 {\bf r}_2) \right) \,, \quad  \lambda = \frac{u}{\kappa + u} \,.
\end{align}
In the latter expression $({\bf r}_1 {\bf r}_2)$ denotes the scalar product. We note that we have skipped 
the constant term $-u_0$ on the right-hand-side of eq. \eqref{H}, because it will cancel out in what follows 
due to a proper normalization. Moreover, for the reasons which will be made clear below, we require that $\lambda^2 < 1$ (and is bounded away from zero), which implies that $-\kappa/2 < u < \infty$ and signifies that the interactions between the dissimilar species can be (modestly) repulsive with negative $u$, or attractive with an arbitrary strength. 
% \begin{align}
 %\label{a}
 %H({\bf r}_1|{\bf r}_2) = \frac{\kappa}{2} \left({\bf r}_1^2 + {\bf r}_2^2\right) + U(r) \,.
% \end{align}
 The focal properties of our analysis are the curls of the probability currents, defined as
 \begin{equation}
 \begin{split}
 \label{curl}
s_1({\bf r_1}|{\bf r_2}) &= \nabla_1 \times {\bf j}_1({\bf r_1}|{\bf r_2}) = \frac{\partial}{\partial x_1} j_1^y({\bf r_1}|{\bf r_2}) -  \frac{\partial}{\partial y_1} j_1^x({\bf r_1}|{\bf r_2}) \,, \\
s_2({\bf r_1}|{\bf r_2}) &=  \nabla_2 \times  {\bf j}_2({\bf r_1}|{\bf r_2}) = \frac{\partial}{\partial x_2} j_2^y({\bf r_1}|{\bf r_2}) -  \frac{\partial}{\partial y_2} j_2^x({\bf r_1}|{\bf r_2}) \,.
 \end{split}
 \end{equation}
 n thermal equilibrium, i.e., for $T_1 = T_2 = T$, the solution of eq. \eqref{fp} is the Boltzmann distribution $P({\bf r_1}|{\bf r_2}) \sim \exp\left(- H({\bf r}_1|{\bf r}_2)/T\right)$, the probability currents in eqs. \eqref{j} and hence, their curls are identically equal to zero. For $T_1 \neq T_2$ this is no longer the case. 
  
Equations \eqref{langevin} can be appropriately generalized to describe the evolution of  a more complex system with $N$ particles of type $1$, and $N$ particles of type $2$, in which the dissimilar particles interact between themselves,  all-with-all via the potential in eq. \eqref{potent},  while the similar species are mutually non-interacting. Labeling the particles $1$ by the index $\alpha$, $\alpha=1,2, \ldots, N$, and the particles $2$ - by the index $\beta$, $\beta = 1,2, \ldots,N$, we denote their respective positions by ${\bf r}_{1,\alpha}$ and ${\bf r}_{2,\beta}$. Then, the Hamiltonian of such a multi-particle system reads
\begin{equation}
\begin{split}
\label{Hmany}
 H(\left\{{\bf r}_{1,\alpha}\right\}|\left\{{\bf r}_{2,\beta}\right\}) &= \left(\kappa + Nu\right) \left(\frac{1}{2}\sum_{\alpha} {\bf r}_{1,\alpha}^2 + \frac{1}{2} \sum_{\beta} {\bf r}_{2,\beta}^2 - \lambda_N \sum_{\alpha} \sum_{\beta} ({\bf r}_{1,\alpha} \cdot {\bf r}_{2,\beta}) \right)  \,,  \\
 \lambda_N &= \frac{u}{\kappa+Nu} \,,
 \end{split}
\end{equation}
where $\alpha,\beta = 1,2, \ldots, N$. Note that 
we have dropped in eq. \eqref{Hmany} (an irrelevant) constant term $\sim u_0$ and also stipulate that $\lambda_N^2 < 1$ implying  that $u$ obeys a double-sided inequality $-\kappa/(2N) < u < \infty$.  
%\begin{align}
%\label{potmany}
%H(\left\{{\bf r}_{1,\alpha}\right\}|\left\{{\bf r}_{2,\beta}\right\}) = \frac{\kappa}{2} \sum_{\alpha} {\bf r}_{1,\alpha}^2 + \frac{\kappa}{2} \sum_{\beta} {\bf r}_{2,\beta}^2 + \sum_{\alpha} \sum_{\beta} U\left(|{\bf r}_{1,\alpha} - {\bf r}_{2,\beta}|\right)
%\end{align} 

In this more general situation the evolution equations become
\begin{equation}
\begin{split}
\dot{\bf r}_{1,\alpha} &=  -  \nabla_{1,\alpha} H(\left\{{\bf r}_{1,\alpha}\right\}|\left\{{\bf r}_{2,\beta}\right\})   + {\bf \zeta}_{1,\alpha} \,, \quad \alpha=1,2, \ldots, N \,, \\
\dot{\bf r}_{2,\beta} &=  -  \nabla_{2,\beta} H(\left\{{\bf r}_{1,\alpha}\right\}|\left\{{\bf r}_{2,\beta}\right\})  + {\bf \zeta}_{2,\beta} \,, \quad \beta=1,2, \ldots, N \,,
\end{split}
\end{equation}
while the multi-particle Fokker-Planck equation in the steady-state attains the form
\begin{equation}
\begin{split}
\label{FPmany}
&0=\sum_{\alpha} \nabla_{1,\alpha} \Big[T_1 \nabla_{1,\alpha} P(\left\{{\bf r}_{1,\alpha}\right\}|\left\{{\bf r}_{2,\beta}\right\}) + P(\left\{{\bf r}_{1,\alpha}\right\}|\left\{{\bf r}_{2,\beta}\right\}) \nabla_{1,\alpha} H(\left\{{\bf r}_{1,\alpha}\right\}|\left\{{\bf r}_{2,\beta}\right\})\Big] \\
&+\sum_{\beta} \nabla_{2,\beta} \Big[T_2 \nabla_{1,\beta} P(\left\{{\bf r}_{1,\alpha}\right\}|\left\{{\bf r}_{2,\beta}\right\}) + P(\left\{{\bf r}_{1,\alpha}\right\}|\left\{{\bf r}_{2,\beta}\right\}) \nabla_{2,\beta} H(\left\{{\bf r}_{1,\alpha}\right\}|\left\{{\bf r}_{2,\beta}\right\})\Big]  \,,
\end{split}
\end{equation}
in which the terms in the brackets are the probability currents, taken with the sign $"-"$.
For such a model, we will also analyze the probability currents and their curls.

\section{Two-particle case}
\label{sec:2particles}

Consider first the behavior of a system with just two particles, which can be also considered as  
%Our analytical approach is based on two simplifying assumptions.  First, we assume that $T_1$ and $T_2$ are both small, $T_{1,2} \ll 1$, and not very different from each other, such that the difference $T_2 - T_1$ is a small parameter. 
%Second, because both temperatures are small, we expect that due to the presence of an optical trap 
%both particles reside close to the bottom of the well and do not travel far away from the origin, such that ${\bf r}_{1,2}$ themselves as well as their difference $r = |{\bf r}_1 - {\bf r}_2|$ are small. We assume next that in this limit the potential $U(r)$ can be approximated by a quadratic function of the form
%\begin{align}
%\label{U}
%U(r) = - u_0  + u \frac{r^2}{2} \,,
%\end{align}
%where $u_0 > 0$ defines the depth of the well at $r =0$, and $u > 0$ is a parameter. 
%We note parenthetically that the model in which
%the interaction potential in eq. \eqref{U} is valid for all inter-particle distances $r$ represents 
a two-dimensional version of the two-bead model studied in \cite{fakhri1,fakhri2}, in a particular geometric setting when both beads connected by a harmonic spring and living at two different temperatures, are both tethered by the respective harmonic springs to the same point, while in the model considered in \cite{fakhri1,fakhri2} the particle are tethered to two arbitrary points $A$ and $B$. 
%Below we present an exact solution of the model with the potential in eq. \eqref{U}, which is valid for arbitrary $T _1$, $T_2$, $u$ and $\kappa$. For potentials which have the form in eq. \eqref{U} for small values of $r$ only, our solution will describe the behavior in the limit of small values of the temperatures. 

For the Hamiltonian in eq. \eqref{H}, it is natural to seek the solution of eq. \eqref{fp} in the form
\begin{align}
\label{P}
P({\bf r}_1|{\bf r}_2) = Z^{-1} \exp\left(- \frac{A}{2}  {\bf r}_1^2 - \frac{B}{2}  {\bf r}_2^2 + \lambda \, C \, ({\bf r}_1 {\bf r}_2) \right) \,,
\end{align}
where $Z$ is the normalization constant,
\begin{align}
Z= \int \int d{\bf r}_1 d{\bf r}_2  \exp\left(- \frac{A}{2}  {\bf r}_1^2 - \frac{B}{2}  {\bf r}_2^2 + \lambda \, C \, ({\bf r}_1 {\bf r}_2) \right) = \frac{4 \pi^2 }{ A B - \lambda^2 C^2} \,.\label{eq:parti}
\end{align}
We insert next the expression \eqref{P} into eqs. \eqref{j} and then plug the resulting expressions  into eq. \eqref{fp}. In doing so,  
 we find by equating the coefficients appearing in front of the spatial variables to zero,  
 that for the potential in eq. \eqref{potent}, arbitrary positive $\kappa$ and $u$ which obeys the inequality $-\kappa/2 < u <  \infty$, the coefficients $A$, $B$ and $C$ are given explicitly by (see also Appendix \ref{sec:appA} for more details)
\begin{equation}
\begin{split}
\label{coefts}
A &=\frac{2 (\kappa + u) \left(2 T_2 + \lambda^2 (T_1 - T_2)\right)}{4 T_1 T_2 + \lambda^2 \left(T_2 - T_1\right)^2} \,, \\
B &=\frac{2 (\kappa + u) \left(2 T_1 + \lambda^2 (T_2 - T_1)\right)}{4 T_1 T_2 + \lambda^2 \left(T_2 - T_1\right)^2} \,, \\
C &= \frac{2 (\kappa + u) \left(T_1 + T_2\right)}{4 T_1 T_2 + \lambda^2 \left(T_2 - T_1\right)^2} \,,
\end{split}
\end{equation}
and
\begin{align}\label{eq:Z1}
Z = \frac{\pi^2 \left(4 T_1 T_2 + \lambda^2 \left(T_2 - T_1\right)^2\right)}{(\kappa + u)^2 (1 - \lambda^2)} \,.
\end{align}

\begin{figure}
	\begin{center}
		\includegraphics[width=70mm]{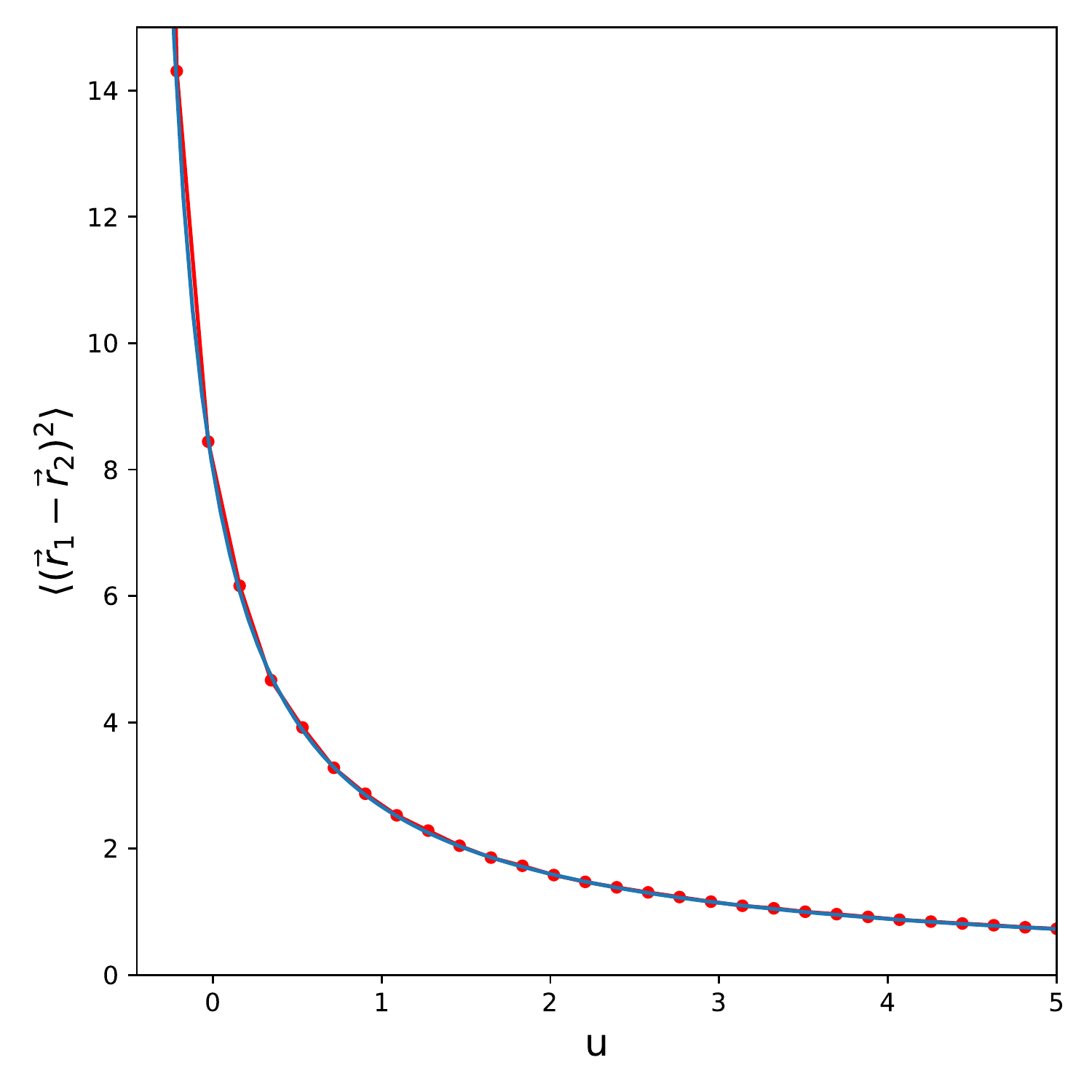}
		\includegraphics[width=70mm]{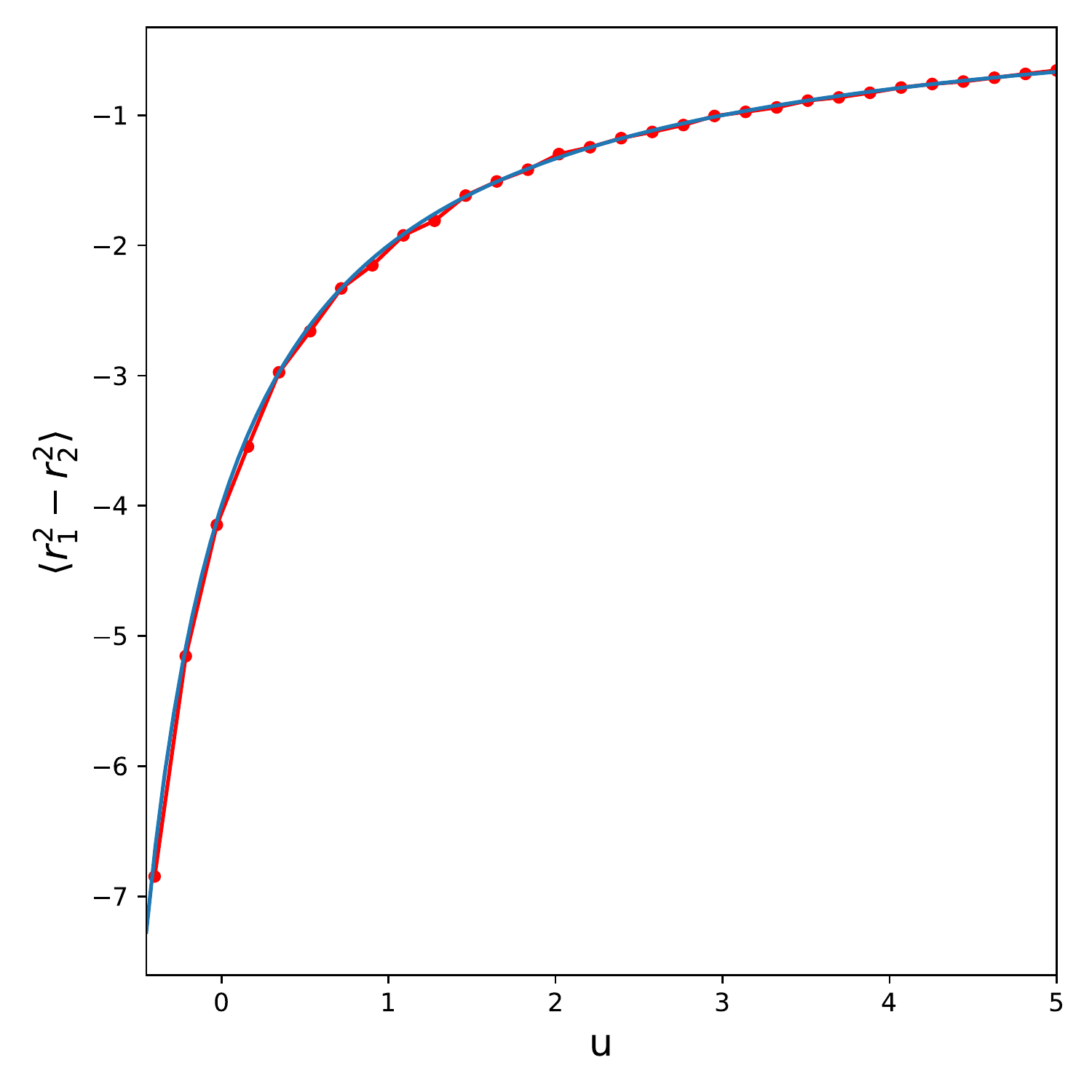}
	\end{center}
	\caption{Panel a:  Mean-squared inter-particle separation distance 
		and panel b:  The difference of mean-squared positions of the two particles  versus the amplitude $u$ of the interaction for $T_1 = 1$ and
		$T_2 = 3$.  Solid curves are defined in eqs. \eqref{eq:reldis}  and \eqref{eq:reldis2}, respectively. Dots present the Monte Carlo simulations results.}
	\label{fig2}
\end{figure}  
In particular, this yields the following simple expression for the mean-squared inter-particle separation distance
\begin{align}
\overline{({\bf r}_1 - {\bf r}_2)^2} &= \int \int d{\bf r}_1 d{\bf r}_2 ({\bf r}_1 - {\bf r}_2)^2 P({\bf r}_1|{\bf r}_2) \nonumber\\&= Z^{-1} \frac{8 \pi^2 (A+B - 2 \lambda C)}{(A B - \lambda^2 C^2)^2} 
= \frac{2 (T_1 + T_2)}{\kappa + 2 u} \,. \label{eq:reldis}
\end{align}
Similarly, one finds that  the difference of the mean-squared displacements of the two particles obeys
\begin{align}
	\overline{{\bf r}_1^2 - {\bf r}_2^2} &= \int \int d{\bf r}_1 d{\bf r}_2 ({\bf r}_1^2 - {\bf r}_2^2) P({\bf r}_1|{\bf r}_2) 
	= \frac{2 (T_1 - T_2)}{\kappa +  u} \,. \label{eq:reldis2}
\end{align}
One infers from eq. \eqref{eq:reldis} that the pair of particles remains {bounded, i.e., $\overline{({\bf r}_1 - {\bf r}_2)^2}$ remains finite,  once $u > - \kappa/2$, which rationalizes the  lower bound on $u$ imposed above.  
Note, as well, that for small $T_1$ and $T_2$ the mean-squared inter-particle separation distance becomes small which signifies that our results may apply in the small-temperature limit to a broader class of the interaction potentials showing a quadratic dependence on $r$ in the limit $r \to 0$. 
Lastly, we observe that $\overline{({\bf r}_1 - {\bf r}_2)^2}$ vanishes when $u \to + \infty$, as one would expect.
%expression \eqref{eq:reldis2} shows that, non counter-intuitively, 
%the particle which is in contact with the hotter thermal bath 
%is located at a distance from the center which is larger than the particle in contact with colder bath.  When the interaction %between particles increases (when $u$ increases), the mean squared distance between particles decreases. 
 \begin{figure}
\begin{center}
\includegraphics[width=70mm]{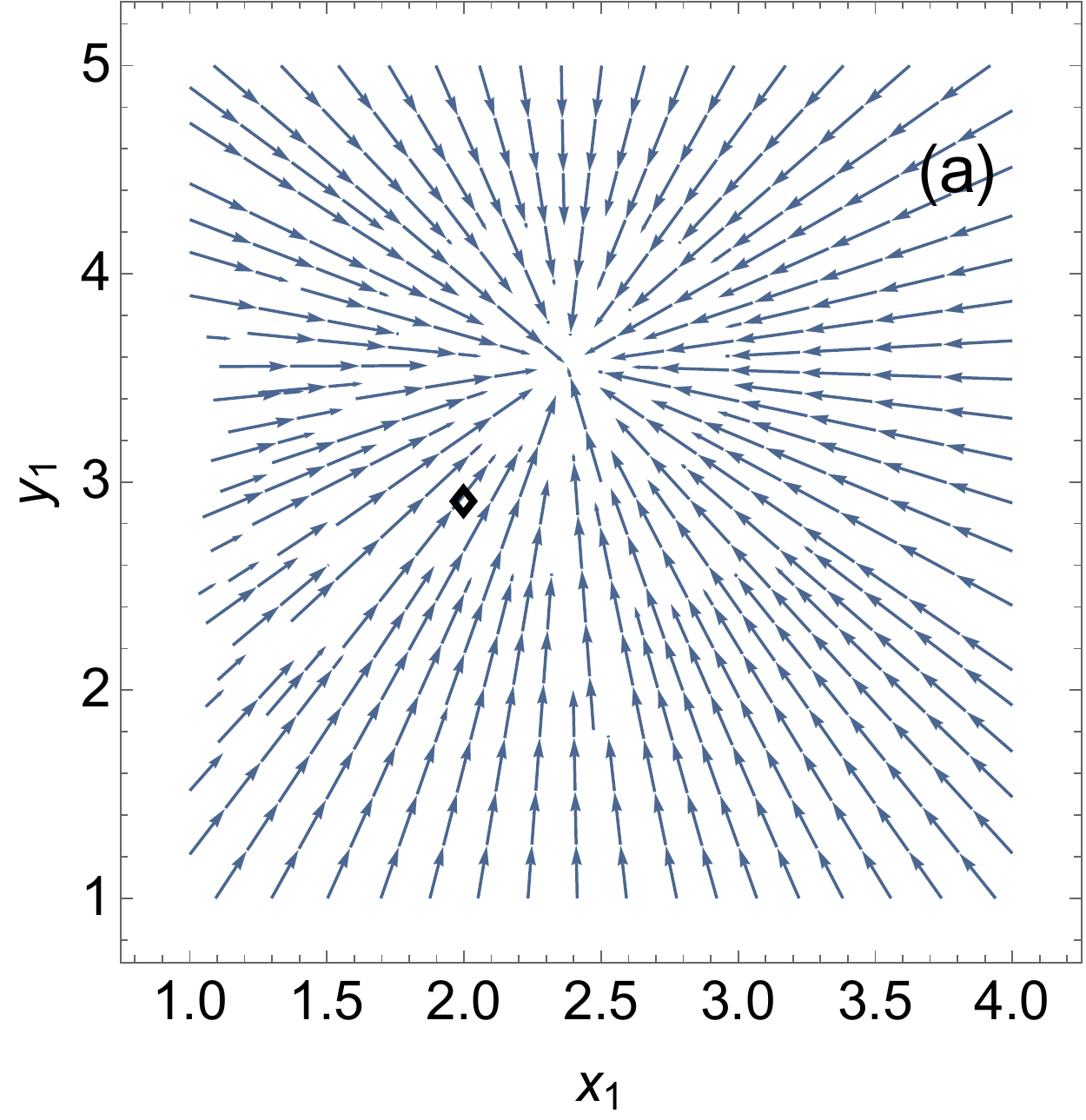}
\includegraphics[width=70mm]{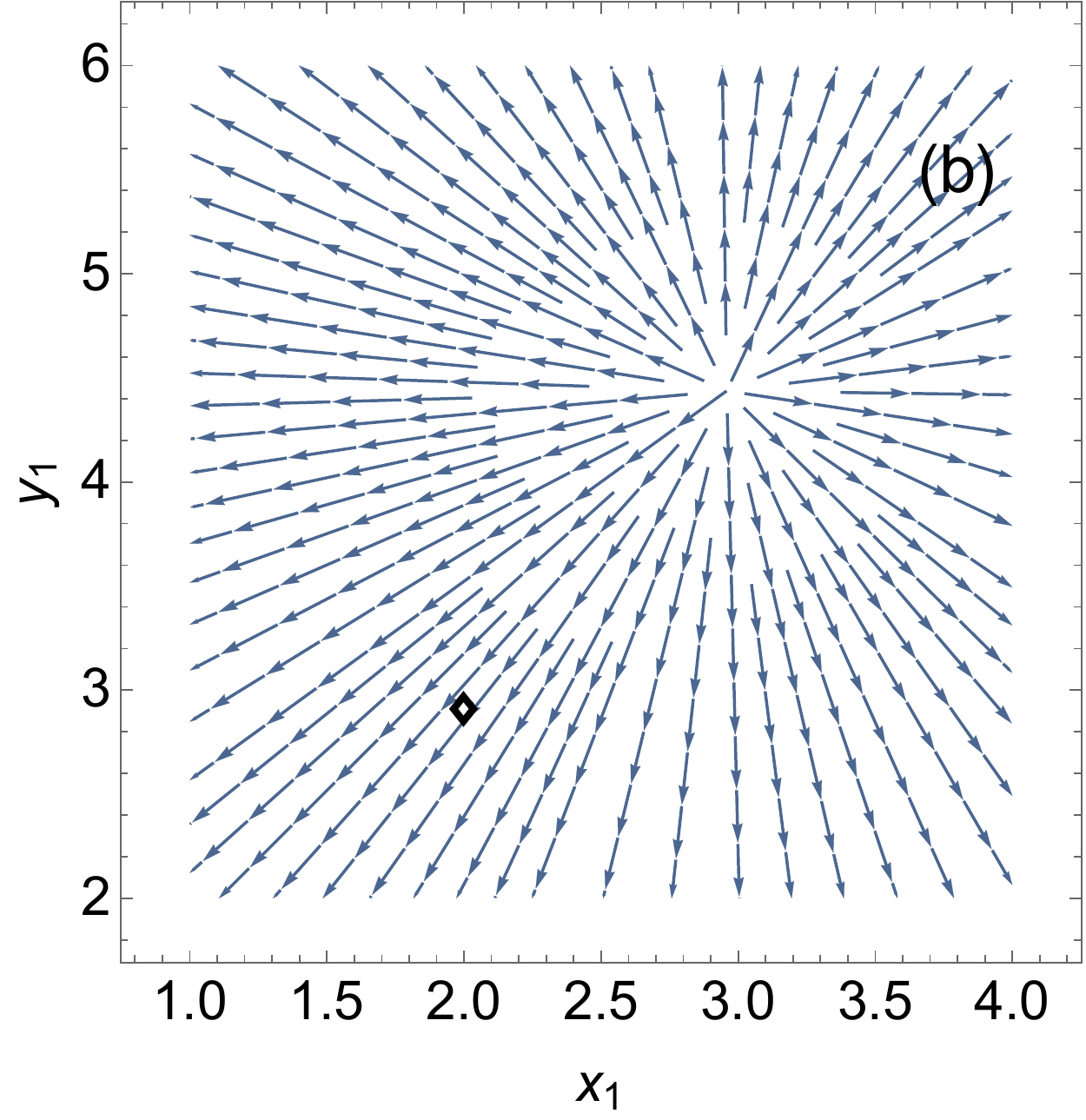}
\includegraphics[width=70mm]{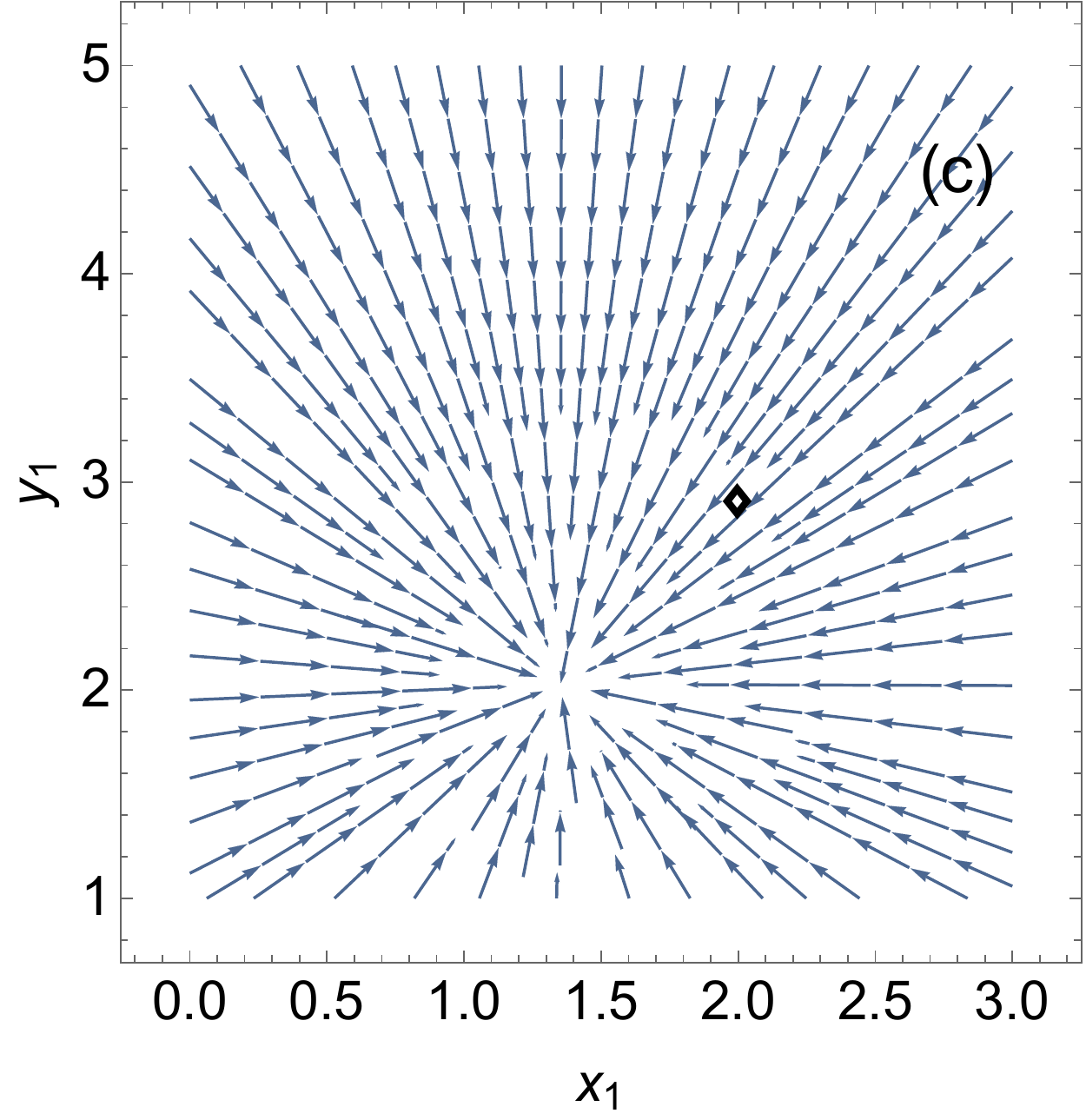}
\includegraphics[width=70mm]{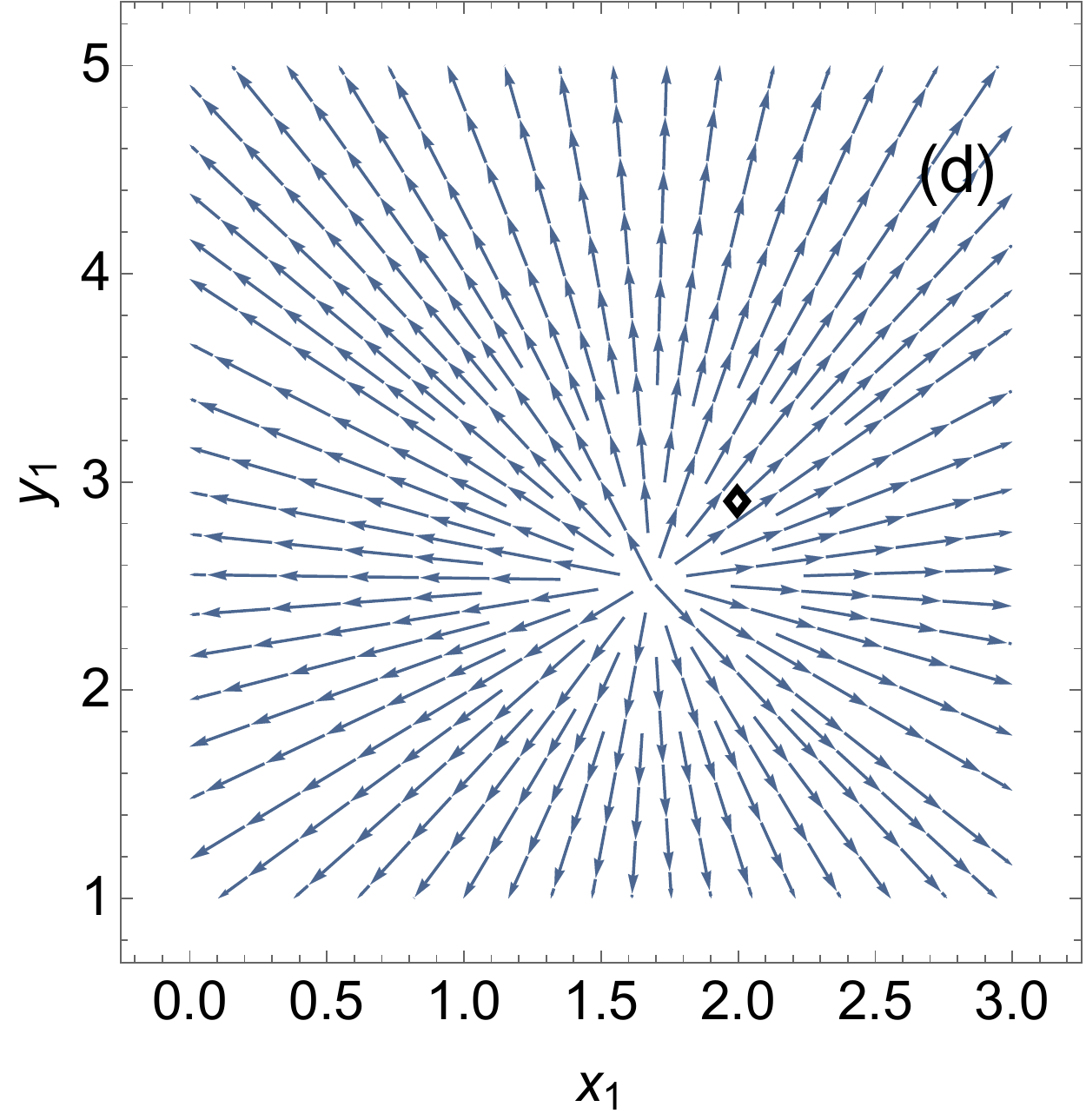}
\end{center}
\caption{Stream plot of the steady-state probability currents ${\bf j}_1({\bf r_1}|{\bf r_2})$ (panels (a) and (b)) and ${\bf j}_2({\bf r_1}|{\bf r_2})$ (panels (c) and (d)) on the $(x_1,y_1)$ plane for a fixed position of the particle $2$ (a diamond),  ${\bf r}_2 = (2,3)$. Parameters $\kappa = 1$ and $u = 3$. Panels (a) and (c): $T_1 = 3$ and
$T_2 = 5$. Panels (b) and (d): $T_1 = 5$ and
$T_2 = 3$. }
\label{fig3}
\end{figure}

We consider next the probability currents defined in eqs. \eqref{j}. Taking advantage of our eq. \eqref{P} together with the  explicit expressions for the coefficients (see eq. \eqref{coefts} above) entering this equation, we have  
\begin{equation}
\begin{split}
\label{jj1}
j_1^x({\bf r}_1|{\bf r}_2) &= \frac{u (T_1 - T_2) \Big[\lambda (T_1+T_2) x_1 - 2 T_1 x_2 + \lambda^2 (T_1 - T_2) x_2
\Big]}{4 T_1 T_2 + \lambda^2 (T_1 - T_2)^2} P({\bf r}_1|{\bf r}_2) \,, \\
j_1^y({\bf r}_1|{\bf r}_2) &= \frac{u (T_1 - T_2) \Big[\lambda (T_1+T_2) y_1 - 2 T_1 y_2 + \lambda^2 (T_1 - T_2) y_2
\Big]}{4 T_1 T_2 + \lambda^2 (T_1 - T_2)^2}  P({\bf r}_1|{\bf r}_2)\,, 
\end{split}
\end{equation}
and
\begin{equation}
\begin{split}
\label{jj2}
j_2^x({\bf r}_1|{\bf r}_2) &=  \frac{u (T_2 - T_1)  \Big[\lambda (T_1+T_2) x_2 - 2 T_2 x_1 + \lambda^2 (T_2 - T_1) x_1
\Big]}{4 T_1 T_2 + \lambda^2 (T_1 - T_2)^2} P({\bf r}_1|{\bf r}_2) \,, \\
j_2^y({\bf r}_1|{\bf r}_2) &=  \frac{u (T_2 - T_1)  \Big[\lambda (T_1+T_2) y_2 - 2 T_2 y_1 + \lambda^2 (T_2 - T_1) y_1
\Big]}{4 T_1 T_2 + \lambda^2 (T_1 - T_2)^2} P({\bf r}_1|{\bf r}_2) \,. 
\end{split}
\end{equation}
Equations \eqref{jj1} and \eqref{jj2} show that the steady-state probability currents vanish in thermal equilibrium, i.e., when $T_1 = T_2$, and also when the parameter $u = 0$, i.e., when the particles $1$ and $2$ are decoupled. For $T_1 \neq T_2$ and $u > 0$, these equations reveal a rather non-trivial 
behavior of the probability currents which 
we depict in Fig.\ref{fig2}. In particular, we observe that for a fixed position of the particle $2$, the currents, e.g.,  $j_1^x({\bf r}_1|{\bf r}_2)$ and $j_1^y({\bf r}_1|{\bf r}_2) $, vanish when the particle $1$ is at position
\begin{equation}
\begin{split}
x_1^* &= \frac{2 T_1 + \lambda^2 (T_2 - T_1)}{\lambda (T_1 + T_2)} x_2 \,, \\
 y_1^* &= \frac{2 T_1 + \lambda^2 (T_2 - T_1)}{\lambda (T_1 + T_2)}  y_2 \,. 
 \end{split}
\end{equation}

Overall, the patterns appear very different from those encountered in the BG model (see \cite{Lam31,Lam32}) which consist of closed elliptic orbits such that the BG performs a rotational motion around the origin. This signifies that the model under study exhibits a completely different behavior in the out-of-equilibrium situations.

 \begin{figure}
\begin{center}
\includegraphics[width=120mm]{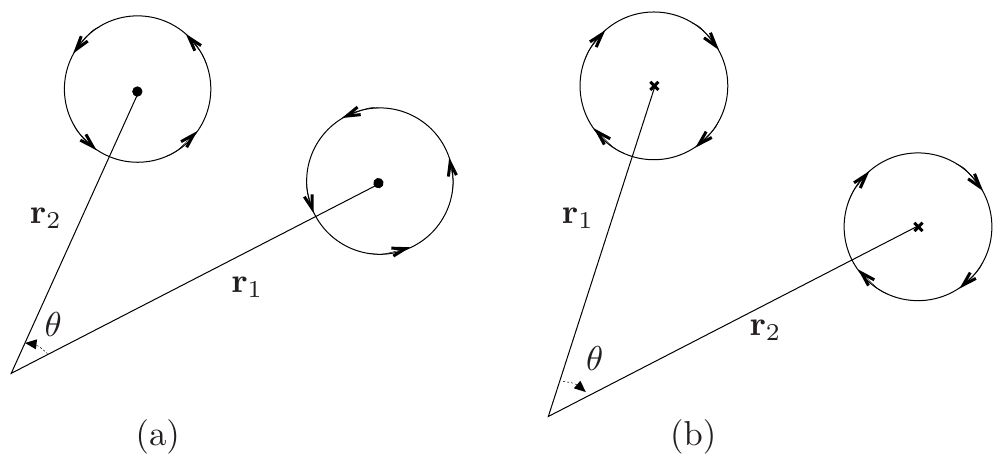}
\end{center}
\caption{Synchronized rotation of the particles $1$ and $2$ for $T_1 \neq T_2$. Both particle rotate clockwise when $\sin(\theta) > 0$ and counter-clockwise when $\sin(\theta) < 0$. No rotation takes place when $\sin(\theta) = 0$ or when either of the particles resides at the origin and also when the positions of both particles obey $x_2 = a x_1$ and $y_2 = a y_1$ with arbitrary value of the scale parameter $a$.}
\label{fig4}
\end{figure}  

\subsection{Spinning tops}

We turn to the most intriguing aspect of the dynamical behavior of the system under study. Namely, we proceed to show that in such a system
 the particles are steadily \textit{spinning} around their centers. 
 This phenomenon can be described in terms of the curl of the probability current -  a vector operator that describes the infinitesimal circulation of a vector field and whose length and direction denote the magnitude and axis of the maximum circulation. Using the definitions of the curls of the probability currents in eqs. \eqref{curl}, we find that the circulation densities $s_1({\bf r}_1|{\bf r}_2)$ and $s_2({\bf r}_1|{\bf r}_2)$ for the particle $1$ occupying position ${\bf r}_1$ and the particle $2$ - position ${\bf r}_2$, obey
 \begin{align}
 \label{s}
s_1({\bf r}_1|{\bf r}_2) & = s_2({\bf r}_1|{\bf r}_2) \nonumber\\
&= \frac{2 u (\kappa + u) (1 - \lambda^2) }{4 T_1 T_2 + \lambda^2 (T_1 - T_2)^2} (T_2 - T_1) \left(x_2 y_1 - x_1 y_2\right) P({\bf r}_1|{\bf r}_2) \,.
\end{align}
We note parenthetically that one obtains exactly 
the expressions \eqref{s} using a definition of the curl in terms of a circular contour integral
\begin{align}
s_1({\bf r}_1|{\bf r}_2)  = \lim_{R \to 0} \frac{1}{\pi R^2} \oint_{|{\bf R|} = R} ds \, {\bf j}_1({\bf r}_1 + {\bf R}|{\bf r}_2) \,.
\end{align}

 \begin{figure}
\begin{center}
\includegraphics[width=65mm]{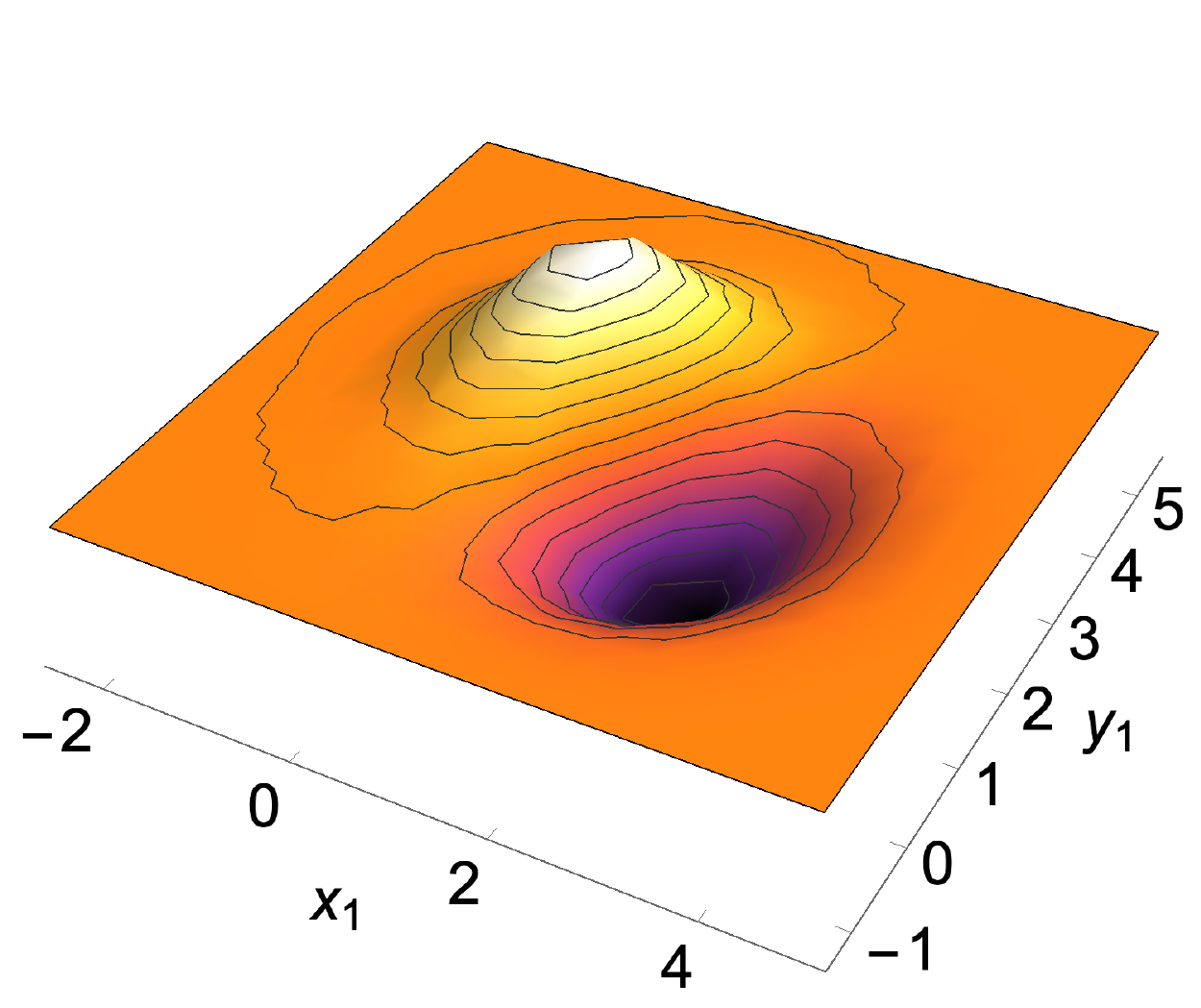}
\includegraphics[width=65mm]{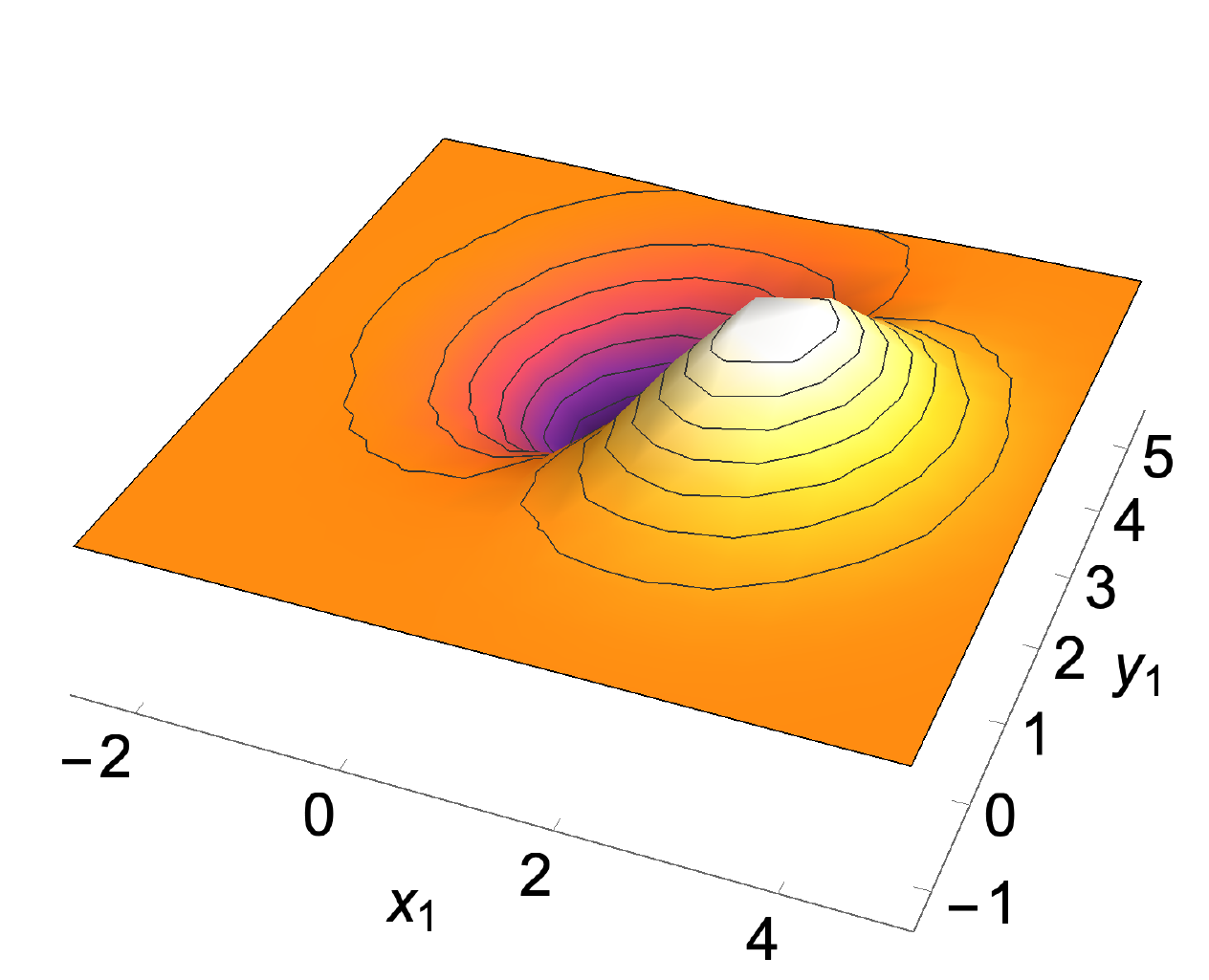}
\end{center}
\caption{Plot of the curl $s_1({\bf r}_1|{\bf r}_2)$ in eq. \eqref{s} for the particle $2$ fixed at position ${\bf r}_2 = (2,3)$ as function of the position of the particle $1$. Parameters $\kappa = 1$ and $u = 3$. Left panel: $T_1 = 3$ and $T_2=5$. Right panel: $T_1 = 5$ and $T_2=3$. 
}
\label{fig5}
\end{figure}  

Expression \eqref{s} reveals several interesting features: \\
(a) Likewise the probability currents, the curl vanishes in thermal equilibrium, i.e. when $T_1 = T_2$.\\
(b) The curl vanishes: (i) when $\kappa \to 0$ ($\lambda \to 1$), i.e., the stiffness of a spring characterizing the optical trap vanishes, (ii) when $u \to - \kappa/2$ ($\lambda \to -1$), i.e., the particles $1$ and $2$ become delocalised and the mean-squared distance between their positions diverges (see eq. \eqref{eq:reldis2}), 
 or (iii) when $u \to 0$, i.e., the particles $1$ and $2$ become decoupled.\\
 (c) For any fixed ${\bf r_1}$ and ${\bf r}_2$, the probability density function $P({\bf r}_1|{\bf r}_2)$ vanishes exponentially when $\kappa$ or $u$ tend to infinity. Given the observation (b), the curl $s_1({\bf r}_1|{\bf r}_2)$ is a non-monotonic function of $\kappa$ and $u$ such that there exist such values of these parameters which provide a maximal (by absolute value) value of the curl.  \\
(d) The curls of the probability currents of the two particles are identically equal to each other for arbitrary positions ${\bf r}_1$ and ${\bf r}_2$, which means that the spinning motion of the particle $1$ and $2$ is always completely synchronized  when only two particles are present in the system.\\
(e) The values of the curls of two particles are position-dependent and proportional to vector product of ${\bf r}_1$ and ${\bf r}_2$. As a consequence, not only the frequency of rotations but also its direction depend on the mutual orientation of the particles $1$ and $2$. In other words, if the sine of the angle between ${\bf r}_1$ and ${\bf r}_2$ is positive, the particles rotate clockwise,  and anti-clockwise, otherwise. Because  
the vector product is anti-commutative, a mere  
 interchange of particles' positions
(see Fig. \ref{fig2}) changes the direction of rotations. \\
(f) The curl vanishes when either of the particles appears at the origin, and also when the positions of both particles obey $x_2 = a x_1$ and $y_2 = a y_1$ with arbitrary scale parameter $a$.\\
(g) Not counter-intuitively, the curl vanishes as $x_1 \to \pm \infty$ or $y_1 \to \pm \infty$
and $(x_2,y_2)$ is fixed, implying that the particles $1$ and $2$ appearing at an infinite separation do not spin.\\
(h) As we observe in Fig. \ref{fig3}, the curl $s_1({\bf r}_1|{\bf r}_2)$ is a non-monotonic function of the position ${\bf r}_1 =(x_1,y_1)$ of the particle $1$ for a fixed position of the particle $2$, ${\bf r}_2 = (x_2,y_2)$ - in fact, it has pronounced maximum and minimum. The extrema at which the curl attains a maximal absolute value, for fixed $x_2$ and $y_2$, are
  $x^{**}_1$ and $y^{**}_1$, which are given by rather cumbersome expressions.  A maximal rotation thus takes place when the particle $1$ is at position $(x_1^{**},y_1^{**})$ and the particle $2$ is at $(x_2,y_2)$. Interestingly enough, such a maximal spinning occurs when the particles are at some finite separation from each other. \\
(i) The curl integrated over all possible positions of one of the particles vanishes,
\begin{align}
 \int d{\bf r}_{1,2} \, s_{1,2}({\bf r}_1|{\bf r}_2) = 0 \,. 
\end{align}

It might be expedient for our further analysis to define a property which does not vanish upon an integration over positions 
of the particle $1$ and particle $2$. Clearly, such a property can be indicative of the emergence of a 
spinning motion for other types of the interaction potentials for which the problem can not be solved and therefore, will be useful for a numerical analysis. To this end, we consider an integrated absolute value of the curl:
\begin{align}
|s_1| = |s_2| = \frac{2 \left(\kappa + u\right) (1 - \lambda^2)  |u (T_1 - T_2)| }{4 T_1 T_2 + \lambda^2 \left(T_1 - T_2\right)^2}\int \int d{\bf r}_{1} d{\bf r}_{2} \, |x_2 y_1 - x_1 y_2| \, P({\bf r}_1|{\bf r}_2) \,.
\end{align}
The four-fold integral in the right-hand-side of the latter equation can be conveniently performed by using 
the integral identity
\begin{align}
|t| = \frac{1}{\pi} \int^{\infty}_{-\infty} \frac{d\omega}{\omega^2} \left(1 - e^{i \omega t} \right) \,,
\label{eq:modt}
\end{align}
which yields
\begin{align}
|s_1| = |s_2| =  |u (T_1 - T_2)| \sqrt{\frac{\left(1 - \lambda^2\right)}{4 T_1 T_2 + \lambda^2 \left(T_1 - T_2\right)^2}} \,.\label{eq:s1}
\end{align}
This property vanishes when $T_1 = T_2$,  when $u \to 0$ and also when $\lambda \to 1$ ($\kappa \to 0$ or $u \to \infty$) or when $\lambda \to - 1$ ($u \to - \kappa/2$).

In Fig. \ref{fig:s12} we depict the dependence of $|s_{1}|$ on $u$ for $\kappa=1$, $T_1= 1$ and $T_2=3$.
The dots present the Monte Carlo simulation data whereas the full curve corresponds  to our theoretical prediction in eq.\eqref{eq:s1}, showing a perfect agreement between our analytic and numerical results.
 \begin{figure}
	\begin{center}
		\includegraphics[width=90mm]{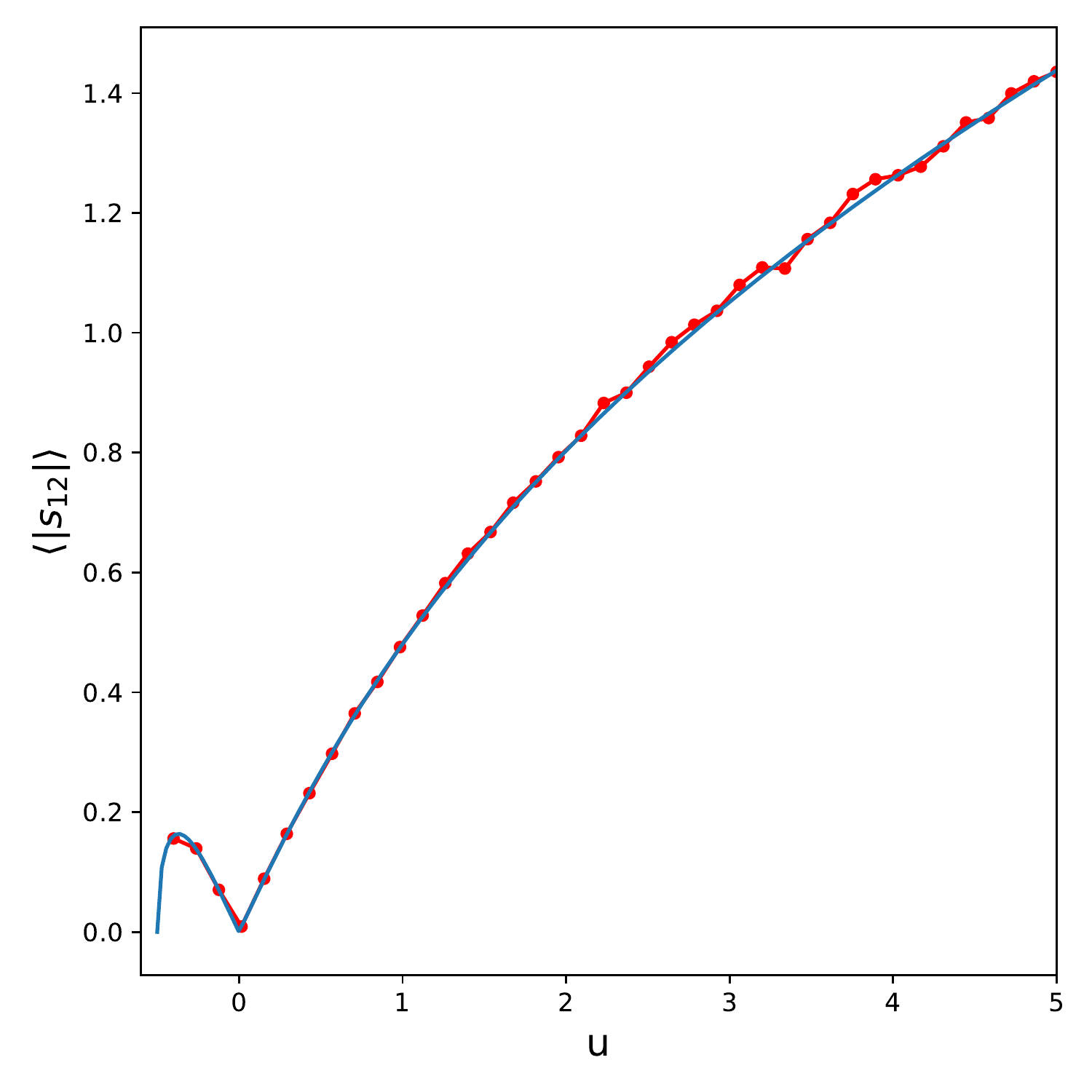}
			\end{center}
	\caption{The integrated absolute value $|s_{1,2}|$ of the curl as a function of $u$ for $T_1=1$ and $T_2=3$. Full curve is the analytical expression (see eq.\eqref{eq:s1}) and dots correspond to the simulation results. 
	}
	\label{fig:s12}
\end{figure} 
 
\subsection{ Correlations of currents}

We pursue our analysis of the model with just two particles in order to show that the spinning of the particles does not provide an exhaustive picture of the dynamical behaviour, which appears to be somewhat richer. To this end, we focus on the behaviour in $4$-dimensional space $(x_1,x_2,y_1,y_2)$ and present the stream plots of the components of the probability in planes $(x_1,x_2)$, $(y_1,y_2)$, $(x_1,y_2)$  and $(x_2,y_1)$ (see Fig. \ref{fig6}). 

 \begin{figure}
\begin{center}
\includegraphics[width=70mm]{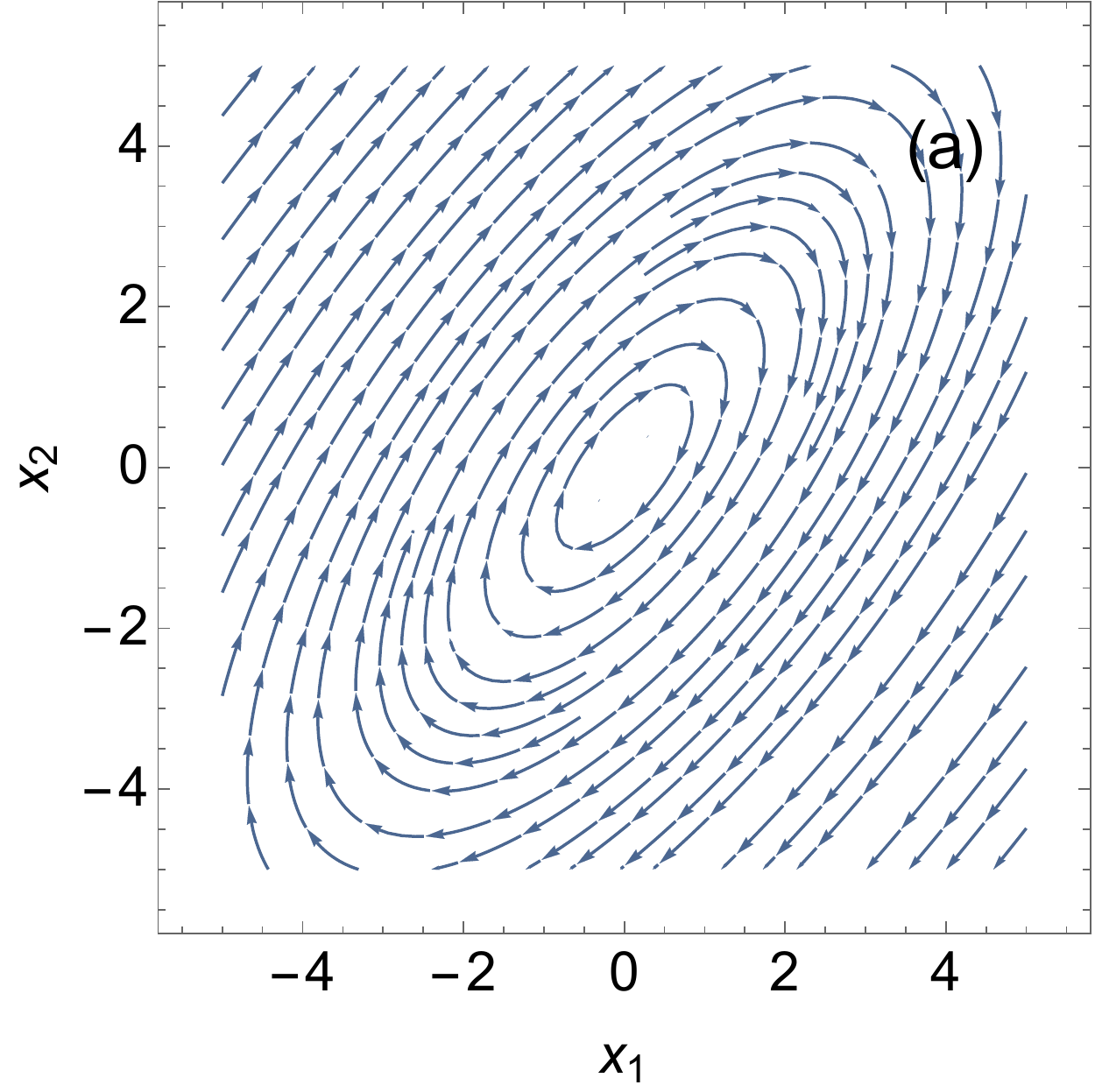}
\includegraphics[width=70mm]{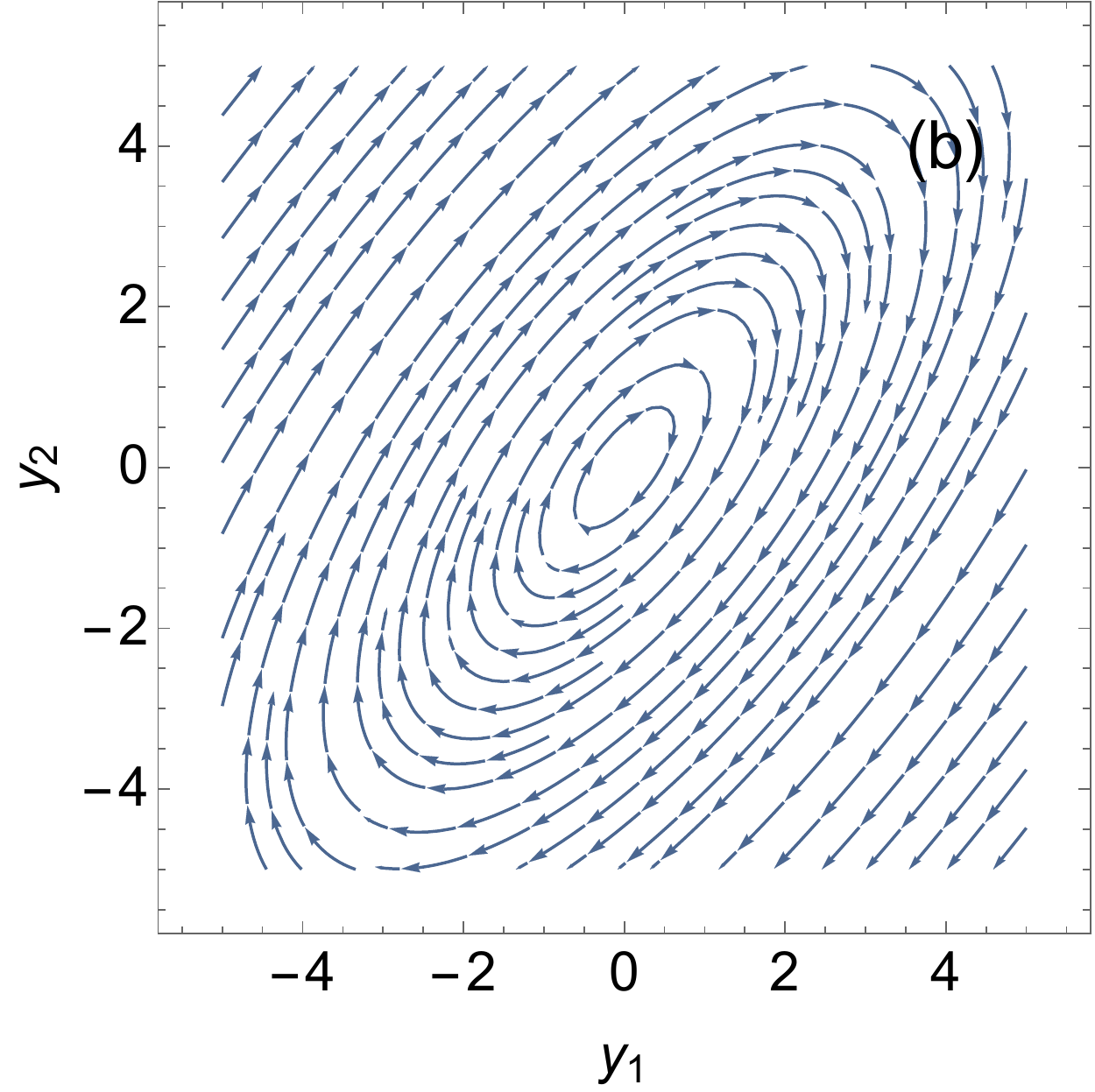}
\includegraphics[width=70mm]{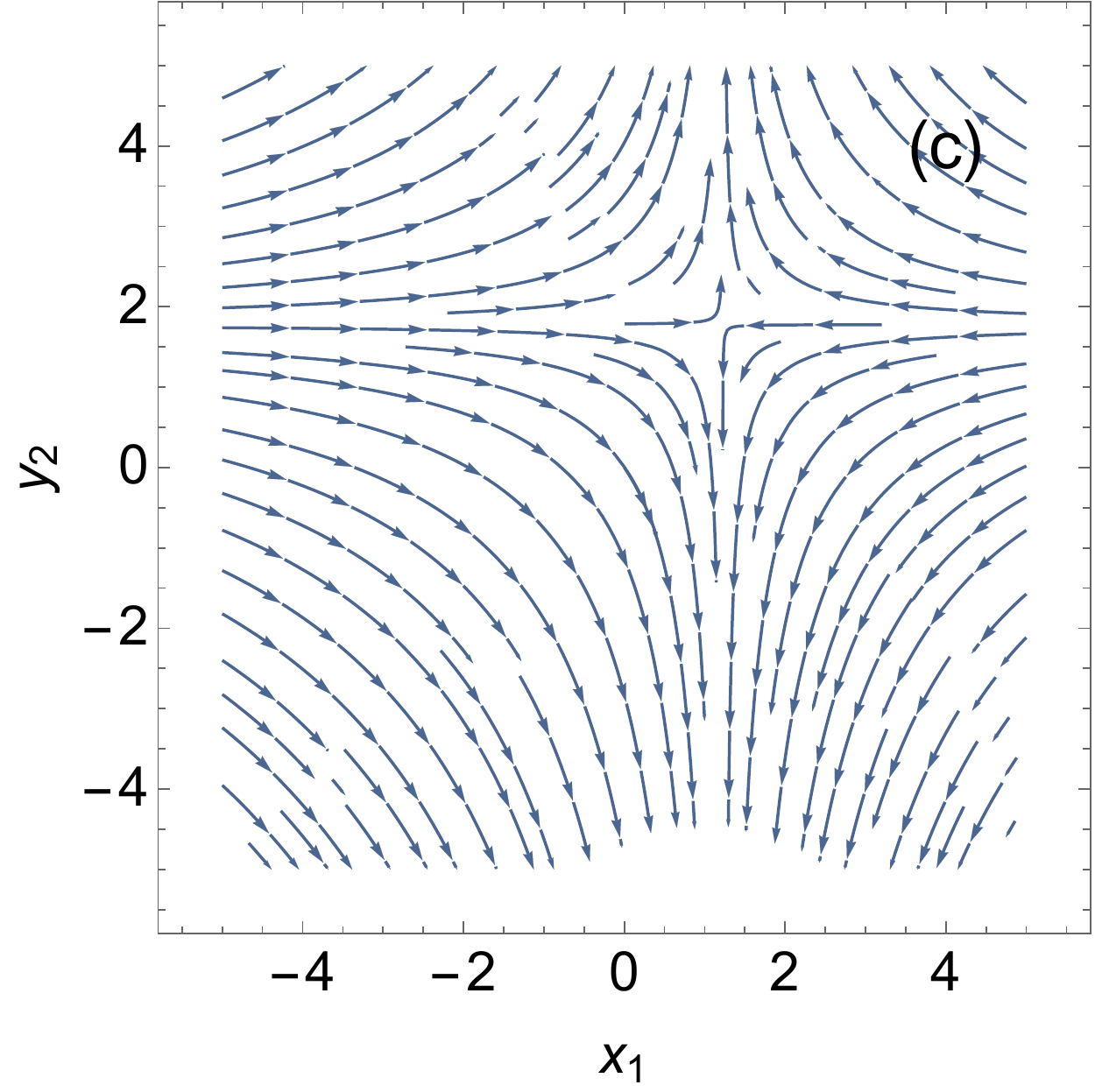}
\includegraphics[width=70mm]{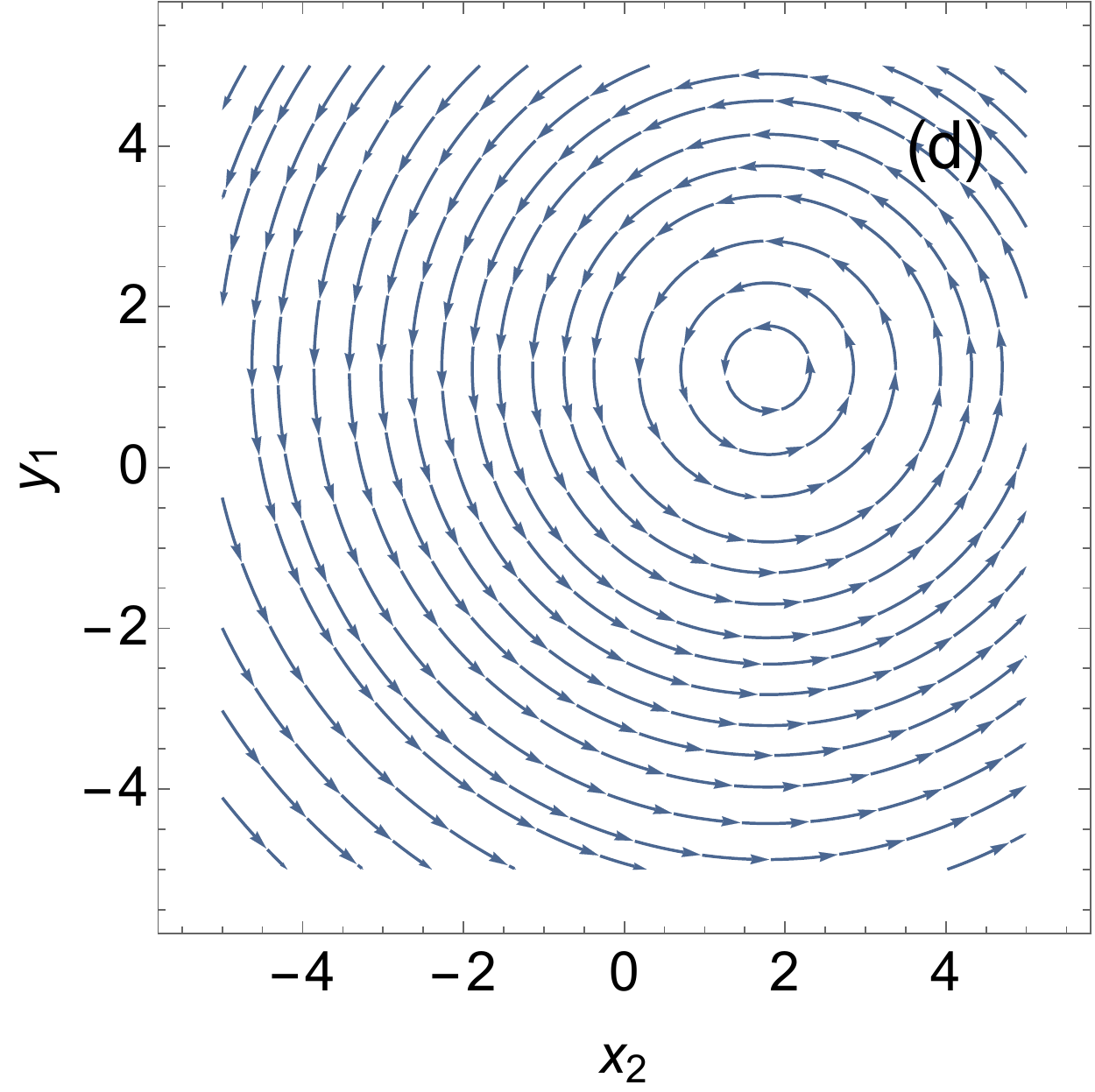}
\end{center}
\caption{Stream plot of the components of the steady-state probability currents in eqs. \eqref{jj1} and \eqref{jj2} in $4$-dimensional space $(x_1,x_2,y_1,y_2)$ for $T_1=1$, $T_2 = 2$, $\kappa = 1$ and $u=2$. Panel (a): Vector with components $(j_1^x,j_2^x)$ 
on the $(x_1,x_2)$-plane for $y_1=y_2=0$. Panel (b): Vector with components $(j_1^y,j_2^y)$ 
on the $(y_1,y_2)$-plane for $x_1=x_2=0$. 
Panel (c):  Vector with components $(j_1^x,j_2^y)$ 
on the $(x_1,y_2)$-plane for $x_2=y_1=1$. Panel (d): Vector with components $(j_1^y,j_2^x)$ 
on the $(x_2,y_2)$-plane for $x_1=y_2=1$. 
}
\label{fig6}
\end{figure}  

 We observe from Fig. \ref{fig6}, (see panels (a) and (b)), that the vectors ${\bf j}^x = (j_1^x,j_2^x)$ and ${\bf j}^y = (j_1^y,j_2^y)$
are circulating along the elliptic orbits, centered at the origin, on the $(x_1,x_2)$- and $(y_1,y_2)$-planes, respectively. Such a behaviour is reminiscent of the one observed previously for the BG model, in which 
the vector $(j_1^x,j_1^y)$ was found to circulate along closed orbits prompting an overall rotational motion of the particle around the origin (see, e.g.,  \cite{Lam31,Lam32}). As evidenced in Fig. \ref{fig3}, this kind of a behaviour is absent here and neither of the particles nor their center of mass are performing a rotational motion around the origin. Here, one may rather claim that the behaviour depicted in Fig. \ref{fig6} mirrors some emergent correlations in the dynamics of the components in $4$-dimensional space. Further on, in panel (c) we present the stream plot of the vector $(j_1^x,j_2^y)$ which shows the $x$-component of the current associated with the particle $1$ and the $y$-component of the current associated with the particle $2$  are anti-correlated and here the pattern consists of parabolas, which travel to $+ \infty$ in the upper plane and $- \infty$ - in the lower plane. Lastly, we find that for the vector $(j_1^y,j_2^x)$ on the $(x_2,y_2)$-plane  the orbits are again closed and form perfect circles
(see panel (d)).

\section{Multi-particle case}\label{sec:Npart}
%\subsection{General results}
%\label{sec:GenNpart}

Consider next a general situation with a population of $N$ particles $1$, living at the temperature $T_1$, and $N$ particles $2$ connected to a heat bath maintained at temperature $T_2$. 
For the Hamiltonian \eqref{Hmany} the solution of the Fokker-Planck equation \eqref{FPmany}  is given explicitly by (details of the derivation can be found in Appendix \ref{sec:appA})
\begin{equation}
\begin{split}
		\label{eq:PPt}
		P(\left\{{\bf r}_{1,\alpha}\right\}|\left\{{\bf r}_{2,\beta}\right\}) &= Z_N^{-1} \exp\Bigg[- \frac{\kappa + N u}{2} \Bigg\{\left(\frac{1}{T_1}  +  \frac{ \lambda_N^2 N \left(T_1^2 - T_2^2\right)}{T_1 \left(4 T_1 T_2 + \lambda_N^2 N^2 \left(T_1 - T_2\right)^2\right)}\right) \sum_{\alpha} {\bf r}_{1,\alpha}^2  \\
		&+ \frac{ \lambda_N^2 N \left(T_1^2 - T_2^2\right)}{T_1 \left(4 T_1 T_2 + \lambda_N^2 N^2 \left(T_1 - T_2\right)^2\right)}  \sum_{\alpha} \sum_{\alpha',  \alpha' \neq \alpha} ({\bf r}_{1,\alpha}   {\bf r}_{1,\alpha'} )  \\
		&  + \left(\frac{1}{T_2}  +  \frac{\lambda_N^2 N \left(T_2^2 - T_1^2\right)}{T_2 \left(4 T_1 T_2 + \lambda_N^2 N^2 \left(T_1 - T_2\right)^2\right)}\right) 
		\sum_{\beta} {\bf r}_{2,\beta}^2 \\& +  \frac{\lambda_N^2 N \left(T_2^2 - T_1^2\right)}{T_2 \left(4 T_1 T_2 + \lambda_N^2 N^2 \left(T_1 - T_2\right)^2\right)} \sum_{\beta} \sum_{\beta',  \beta'  \neq \beta} ({\bf r}_{2,\beta} {\bf r}_{2,\beta'}) \\
		& - \frac{4 \lambda_N \left(T_1 + T_2\right)}{ \left(4 T_1 T_2 + \lambda_N^2 N^2 \left(T_1 - T_2\right)^2\right)} \, \sum_{\alpha} \sum_{\beta} ({\bf r}_{1,\alpha}  {\bf r}_{2,\beta})
		\Bigg\}\Bigg] \,,
\end{split}
\end{equation}
where $Z_N$ is the normalization constant which is defined in eq. \eqref{eq:norm2}. For $N = 1$, the expression \eqref{eq:PPt} evidently reduces to the result in eqs. \eqref{P} to \eqref{coefts}.  
All moments of the multivariate distribution can be then obtained directly from eqs. \eqref{eq:PPt} and \eqref{eq:norm2}. 

Further on, we get the following exact expressions for the probability currents: 
\begin{equation}
\begin{split} 
\label{Ncur1}
		j^x_{1,\alpha}\left(\left\{{\bf r}_{1,\alpha}\right\}|\left\{{\bf r}_{2,\beta}\right\}\right) &= \Bigg[ \frac{ u N \left[2 T_1 (T_2-T_1) + \lambda_N^2 N^2 (T_1 - T_2)^2\right] X_2 }{4 T_1 T_2 + \lambda_N^2 N^2 (T_1 - T_2)^2} + \\
		&+ \frac{u \lambda_N N^2 (T_1^2-T_2^2) X_1}{4 T_1 T_2 + \lambda_N^2 N^2 (T_1 - T_2)^2} \Bigg] P(\left\{{\bf r}_{1,\alpha}\right\}|\left\{{\bf r}_{2,\beta}\right\}) \,, \\
		j^y_{1,\alpha}\left(\left\{{\bf r}_{1,\alpha}\right\}|\left\{{\bf r}_{2,\beta}\right\}\right) &= \Bigg[ \frac{ u N \left[2 T_1 (T_2-T_1) + \lambda_N^2 N^2 (T_1 - T_2)^2\right] Y_2 }{4 T_1 T_2 + \lambda_N^2 N^2 (T_1 - T_2)^2} + \\
		&+ \frac{u \lambda_N N^2 (T_1^2-T_2^2) Y_1}{4 T_1 T_2 + \lambda_N^2 N^2 (T_1 - T_2)^2} 
		\Bigg] P(\left\{{\bf r}_{1,\alpha}\right\}|\left\{{\bf r}_{2,\beta}\right\}) \,,
\end{split}
\end{equation} 
and
\begin{equation}
\begin{split}
\label{Ncur2}
j^x_{2,\beta}\left(\left\{{\bf r}_{1,\alpha}\right\}|\left\{{\bf r}_{2,\beta}\right\}\right) &= \Bigg[ \frac{ u N \left[2 T_2 (T_1-T_2) + \lambda_N^2 N^2 (T_1 - T_2)^2\right] X_1 }{4 T_1 T_2 + \lambda_N^2 N^2 (T_1 - T_2)^2} + \\
		&+ \frac{u \lambda_N N^2 (T_2^2-T_1^2) X_2}{4 T_1 T_2 + \lambda_N^2 N^2 (T_1 - T_2)^2}
\Bigg] P(\left\{{\bf r}_{1,\alpha}\right\}|\left\{{\bf r}_{2,\beta}\right\})  \,,  \\
j^y_{2,\beta}\left(\left\{{\bf r}_{1,\alpha}\right\}|\left\{{\bf r}_{2,\beta}\right\}\right) &= \Bigg[ \frac{ u N \left[2 T_2 (T_1-T_2) + \lambda_N^2 N^2 (T_1 - T_2)^2\right] Y_1 }{4 T_1 T_2 + \lambda_N^2 N^2 (T_1 - T_2)^2} + \\
		&+ \frac{u \lambda_N N^2 (T_2^2-T_1^2) Y_2}{4 T_1 T_2 + \lambda_N^2 N^2 (T_1 - T_2)^2}
\Bigg]  P(\left\{{\bf r}_{1,\alpha}\right\}|\left\{{\bf r}_{2,\beta}\right\})  \,,
\end{split}
\end{equation}
where $X_1$, $Y_1$, $X_2$ and $Y_2$ denote respectively the positions of the centers of mass of two ensembles of particles:
\begin{equation}
\begin{split}
\label{cm}
X_1 &= \frac{1}{N} \sum_{\alpha} x_{1,\alpha} \,, \qquad X_2 = \frac{1}{N} \sum_{\beta} x_{2,\beta} \,, \\
Y_1 &= \frac{1}{N} \sum_{\alpha} y_{1,\alpha} \,, \qquad Y_2 = \frac{1}{N} \sum_{\beta} y_{2,\beta} \,.
\end{split}
\end{equation}
Equations \eqref{Ncur1} and \eqref{Ncur2} generalize eqs. \eqref{jj1} and \eqref{jj2} of the previous section obtained for the case $N=1$. Similarly to the behavior observed in the $N = 1$ case, the probability currents in the general case vanish once the temperatures $T_1$ and $T_2$ are equal to each other.
One notices next that the prefactors in front of the probability density function is again a linear combination of (all) particles' positions and vanish when $X_1$ ($Y_1$) and $X_2$ ($Y_2$) obey certain linear relations. Therefore, the probability currents exhibit essentially the same kind of  a behavior as depicted in Fig. \ref{fig3} for the case $N = 1$.

Finally, we concentrate on the curls of the probability currents given by eqs. \eqref{Ncur1} and \eqref{Ncur2}. We readily find
\begin{equation}
\begin{split}
\label{zu}
&s_{1,\alpha}\left(\left\{{\bf r}_{1,\alpha}\right\}|\left\{{\bf r}_{2,\beta}\right\}\right)   = \frac{\partial}{\partial x_{1,\alpha}} j^y_{1,\alpha}\left(\left\{{\bf r}_{1,\alpha}\right\}|\left\{{\bf r}_{2,\beta}\right\}\right) - \frac{\partial}{\partial y_{1,\alpha}} j^x_{1,\alpha}\left(\left\{{\bf r}_{1,\alpha}\right\}|\left\{{\bf r}_{2,\beta}\right\}\right) \\
&= P(\left\{{\bf r}_{1,\alpha}\right\}|\left\{{\bf r}_{2,\beta}\right\}) \frac{u N (\kappa + N u) (T_1 - T_2)}{T_1 \left(4 T_1 T_2 + \lambda_N^2 N^2 (T_1 - T_2)^2\right)} \Big[\lambda_N^2 N^2 (T_1+T_2) \left(X_2 Y_1 - Y_2 X_1\right) \\&+ \left(2 T_1 - \lambda_N^2 N^2 (T_1 - T_2)\right) \left(x_{1,\alpha} Y_2 - y_{1,\alpha} X_2\right) 
- \lambda_N N (T_1+T_2) \left(x_{1,\alpha} Y_1 - y_{1,\alpha} X_1\right) \Big] 
\end{split}
\end{equation}
and
\begin{equation}
\begin{split}
\label{zu_1}
&s_{2,\beta}\left(\left\{{\bf r}_{1,\alpha}\right\}|\left\{{\bf r}_{2,\beta}\right\}\right)   = \frac{\partial}{\partial x_{2,\beta}} j^y_{2,\beta}\left(\left\{{\bf r}_{1,\alpha}\right\}|\left\{{\bf r}_{2,\beta}\right\}\right) - \frac{\partial}{\partial y_{2,\beta}} j^x_{2,\beta}\left(\left\{{\bf r}_{1,\alpha}\right\}|\left\{{\bf r}_{2,\beta}\right\}\right) \\
& = P(\left\{{\bf r}_{1,\alpha}\right\}|\left\{{\bf r}_{2,\beta}\right\})\frac{u N (\kappa + N u) (T_1- T_2)}{T_2 \left(4 T_1 T_2 + \lambda_N^2 N^2 (T_1 - T_2)^2\right)} \Big[ \lambda_N^2 N^2 (T_1 + T_2) \left(X_2 Y_1 - Y_2 X_1\right) \\
&+ \lambda_N N (T_1+T_2) \left(x_{2,\beta} Y_2 - y_{2,\beta} X_2\right) - \left(2 T_2 + \lambda_N^2 N^2 (T_1 - T_2)\right)  \left(x_{2,\beta} Y_1 - y_{2,\beta} X_1\right) 
\Big]
\end{split}
\end{equation}
%\subsection{Four particle case}
%\label{sec:4part}
%We focus here on the model with 2 particles of species $1$ and $2$ particles
%of species $2$.
The expressions \eqref{zu} and \eqref{zu_1} generalise our result in eq. \eqref{s} over the case $N > 1$ (and reproduce \eqref{s} when we set $N=1$ in eqs. \eqref{zu} and \eqref{zu_1}). 
They are evidently more complicated and do not reveal a complete synchronization of the spinning of particles, as we have observed in the case of just two particles. Here, the curls of the currents depend on the instantaneous positions of all the particles in both ensembles. On the other hand, the curls are clearly not equal to zero, in general, which signifies that 
spinning exists, unless $T_1 = T_2$, $\kappa=0$ or $u=0$. 

A complete synchronization of the spinning motion  becomes apparent, however, if we consider the curls averaged over the corresponding ensembles of particles.  Summing eq. \eqref{zu} over all $\alpha$-s and eq. \eqref{zu_1} over $\beta$-s, we arrive at the following identity:
\begin{equation}
\begin{split}
&\frac{1}{N} \sum_{\alpha} s_{1,\alpha}\left(\left\{{\bf r}_{1,\alpha}\right\}|\left\{{\bf r}_{2,\beta}\right\}\right)= \\
&= \frac{u N (\kappa + N u) (1 - \lambda_N^2 N^2) (T_1 - T_2)}{4 T_1 T_2 + \lambda_N^2 N^2 (T_1 - T_2)^2} \left(X_1 Y_2 - Y_1 X_2\right) P(\left\{{\bf r}_{1,\alpha}\right\}|\left\{{\bf r}_{2,\beta}\right\})\label{s12sum} \\&= \frac{1}{N} \sum_{\beta} s_{2,\beta}\left(\left\{{\bf r}_{2,\beta}\right\}|\left\{{\bf r}_{2,\beta}\right\}\right)
\end{split} 
\end{equation}
%{\alb The term $\lambda_N N\to 1$ for $N\to \infty$, so eq.~\eqref{s12sum} vanishes, unless the term $1- \lambda_N^2 N^2$ simplifies with a similar term. }
Clearly, the same identity holds also when $x_{1,\alpha} = X_1$, $y_{1,\alpha} = Y_1$, $x_{2,\beta} = X_2$ and $y_{2,\beta} = Y_2$, i.e., the positions of two particles coincide with the centers of mass of their respective ensembles.  

\begin{figure}
	\begin{center}
		\includegraphics[width=90mm]{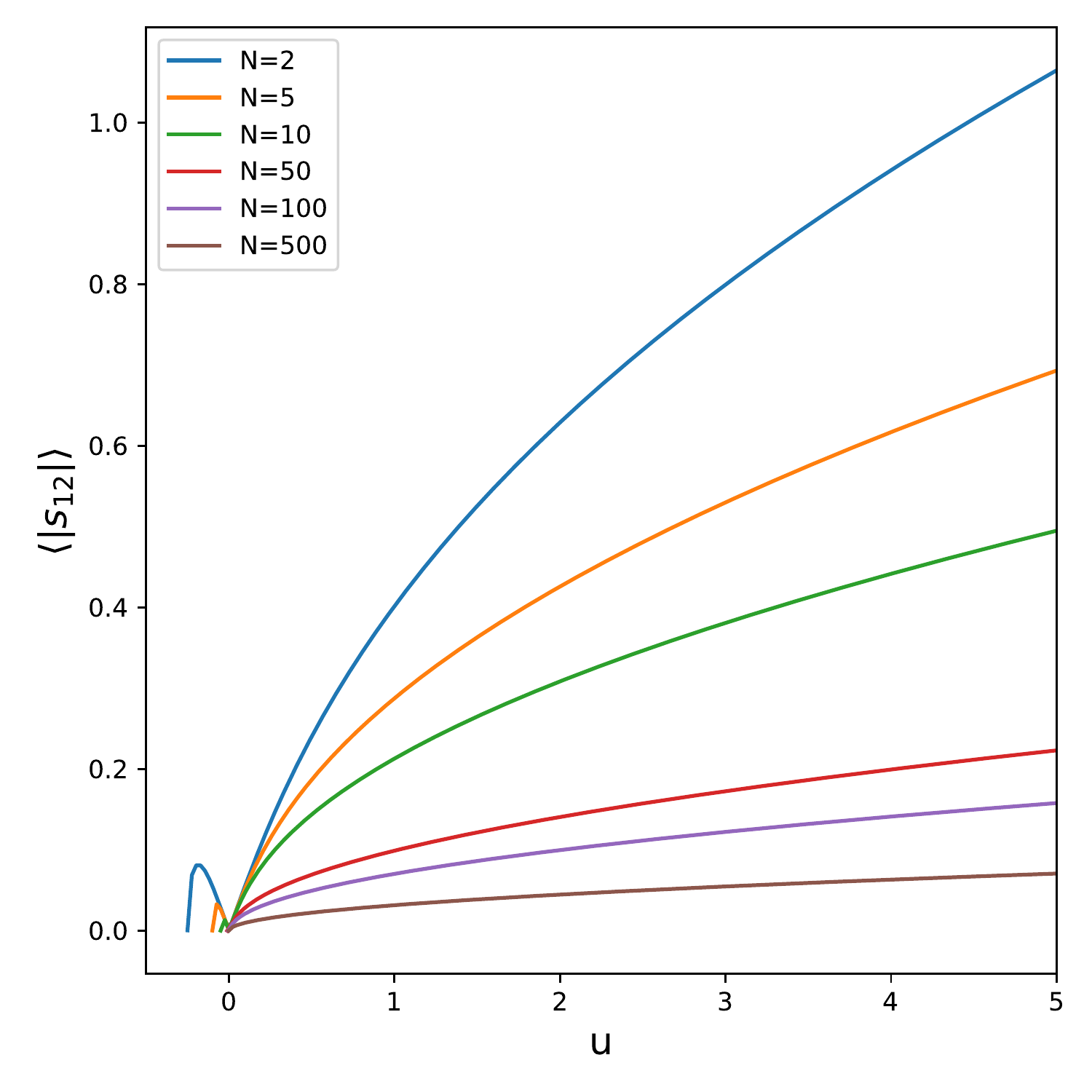}
	\end{center}
	\caption{$|s_{1,2}|$ as a function of $u$ for $T_1=1$ and $T_2=3$. Full curves correspond to  the exact expression (see eq.\eqref{eq:synchro4}) for different values of  $N$.
	}
	\label{fig:sn12}
\end{figure}
For the general case, we also find an averaged absolute value of the curl:
\begin{equation}
\begin{split}
	|s_1| = |s_2|& = \frac{2 N\left(\kappa + N u\right) (1 - \lambda^2_N N ^2)  |u (T_1 - T_2)| }{4 T_1 T_2 + \lambda^2_N N^2 \left(T_1 - T_2\right)^2}\\
&\quad \times \int \int \prod_{\alpha}d{\bf r}_{1,\alpha} \prod_{\beta}d{\bf r}_{2,\beta} \, |X_2 Y_1 - X_1 Y_2| \, P({\bf r}_{1,\alpha}|{\bf r}_{2,\beta}) \\
&=|u(T_1-T_2)|\sqrt{\frac{(1 - \lambda^2_N N ^2)}{4 T_1 T_2 + \lambda^2_N N^2 \left(T_1 - T_2\right)^2}} \,.\label{eq:synchro}
\end{split}
\end{equation}
The details of the  derivation of the explicit expression in the last line of eq. \eqref{eq:synchro}
are presented in Appendix \ref{sec:appA}.

One notices that also in the general case the averaged absolute value of the curl vanishes
when $u \to 0$, as it should, and 
 given the expression for $\lambda_N$ (see  eq.~(\ref{Hmany})), we can also analyze the limiting behavior of $|s_{1,2}|$. We  observe that $|s_{1,2}|$ vanishes, $|s_{1,2}| \propto \sqrt{u + \kappa/(2N)}$,  when $u$ tends to the left extremity of the interval on which it is defined, i.e., $u \to - \kappa/(2N)$.
 Conversely, when $u$ increases, the asymptotic behavior is given by
 \begin{equation}
 	|s_{1,2}|\propto \frac{|(T_1 - T_2)|}{T_1+T_2} \sqrt{\frac{\kappa |u|}{N}},
 \end{equation} 
which means that the collective effect decreases with $N$,  for a large but finite $u$.
 
In regard to the limit $N\to \infty$, it is worth noting that the Hamiltonian \eqref{Hmany} involves long-range interactions between dissimilar particles, such that their contribution 
to the potential energy of the system 
 grows as the squared number of particles, whereas the term associated with 
  the tweezer terms is only extensive. Therefore, in order to make both terms extensive, it might be instructive 
  to turn to the Kac's limit replacing $u$ is by $u/N$. In doing so, we find that the dominant in the limit $N \to \infty$ behavior of $ |s_{1,2}|$ is given by
\begin{equation}
 |s_{1,2}| = \frac{|(T_1 - T_2)u| }{N(T_1+T_2)} \sqrt{\frac{(1 - \lambda^2_1 )}{4 T_1 T_2 + \lambda^2_1  \left(T_1 - T_2\right)^2}} 
\end{equation} 
where $\lambda_1=u/(\kappa+u)$.
As expected, the synchronization effect also vanishes in this limit, but more rapidly with $N$, because the strength of the particle-particle interactions is smaller here than in the previous situation.
 
%corresponding to a system with an infinite number of particles, which is able to sustain a spinning. Note that in this limit $u$ becomes bounded by zero, i.e., is strictly positive.

\section{Conclusion}
\label{conc}

To conclude, we have proposed here a novel minimalist model that exhibits a highly non-trivial behavior under out-of-equilibrium conditions. The model under study comprises two populations of particles, of the same size $N$, and each population is living at its own temperature. All particles move randomly on a two-dimensional plane in the field of a single optical tweezer, i.e., they are all attached to the origin by harmonic springs of a constant strength. Moreover, dissimilar species interact between themselves via a harmonic potential, eq. \eqref{potent}.  Similar species are not mutually interacting.

The model is simple enough to allow an exact solution which shows that in case when the temperatures are not equal to each other, 
there emerge non-zero  probability currents with non-zero position-dependent curls. Physically, it means that the particles performing a Brownian motion in a quadratic potential are steadily spinning around their respective centers of mass, as a kind of "spinning top" toys. Remarkably, the components of the currents associated with the positions of dissimilar particles appear to exhibit non-trivial cooperative effects, such that these components circulate along closed elliptic orbits. 

We admit that the present model can be hardly realized experimentally and hence, our findings here 
can be more of a conceptual 
value. On the other hand, a situation in which each population is held by its \text{own} tweezer 
centered at different points and manipulated differently such that effectively each population lives at its own temperature, has been experimentally realized 
in \cite{alb1}. Moreover, in such a setup  
the coupling between dissimilar species is provided by the hydrodynamic interactions 
resulting in the potential of the form in eq. \eqref{potent}. On intuitive grounds, we expect that the kind 
 of a non-trivial behavior revealed here will be also present in such a situation. This question will be studied elsewhere \cite{alb2}.

\section*{Acknowledgments}

The authors acknowledge helpful discussions with Luca Peliti and Giacomo Gradenigo. AI wishes to thank LPTMC, Sorbonne Universit\'e, for a warm hospitality during his visit in June 2022.

\appendix

\section{Solution in the multi-particle case}
\label{sec:appA} 

We seek the solution $P(\left\{{\bf r}_{1,\alpha}\right\}|\left\{{\bf r}_{2,\beta}\right\})$ of the  Fokker-Planck equation \eqref{FPmany} in the steady-state as an exponential 
of a 
general quadratic form
\begin{align}
	\label{PP}
	P(\left\{{\bf r}_{1,\alpha}\right\}|\left\{{\bf r}_{2,\beta}\right\}) &= Z_N^{-1} \exp\Bigg[- \frac{A_N}{2} \sum_{\alpha} {\bf r}_{1,\alpha}^2 - \frac{B_N}{2} \sum_{\beta} {\bf r}_{2,\beta}^2 + \lambda_N \, C_N \, \sum_{\alpha} \sum_{\beta} ({\bf r}_{1,\alpha} \cdot {\bf r}_{2,\beta}) \nonumber\\ &- \frac{D_N}{2} \sum_{\alpha} \sum_{\alpha',  \alpha' \neq \alpha} ({\bf r}_{1,\alpha} \cdot  {\bf r}_{1,\alpha'} ) - \frac{E_N}{2}  \sum_{\beta} \sum_{\beta',  \beta'  \neq \beta} ({\bf r}_{2,\beta} \cdot {\bf r}_{2,\beta})\Bigg] \,,
\end{align}
where $Z_N$ is the normalization constant, while 
$A_N$, $B_N$, $C_N$, $D_N$ and $E_N$ which are functions of all system's parameters to be determined. Note that for $N = 1$ the sums $\sum_{\alpha} \sum_{\alpha',  \alpha' \neq \alpha} $ and $\sum_{\beta} \sum_{\beta',  \beta'  \neq \beta}$ are formally equal to zero themselves, such that the coefficients $D_N$ and $E_N$ do not have to vanish when $N$ is set equal to $1$. 

Differentiating eq. \eqref{PP},  we find that the components of the currents ${\bf j}_{1,\alpha}$ and ${\bf j}_{2,\beta}$ are explicitly given by
\begin{align}
		j^x_{1,\alpha}\left(\left\{{\bf r}_{1,\alpha}\right\}|\left\{{\bf r}_{2,\beta}\right\}\right) &= \Big[\left(A_N T_1 - D_N T_1  - \kappa - N u\right) x_{1,\alpha} + N \left(u  - \lambda_N C_N T_1\right) X_2+ \nonumber\\
		&+ N D_N T_1 X_1 \Big] P(\left\{{\bf r}_{1,\alpha}\right\}|\left\{{\bf r}_{2,\beta}\right\})  \nonumber\\
		j^y_{1,\alpha}\left(\left\{{\bf r}_{1,\alpha}\right\}|\left\{{\bf r}_{2,\beta}\right\}\right) &= \Big[\left(A_N T_1 - D_N T_1 - \kappa - N u\right) y_{1,\alpha} + N \left(u  - \lambda_N C_N T_1\right) Y_2 + \nonumber\\ &+ N D_N T_1 Y_1 
		\Big] P(\left\{{\bf r}_{1,\alpha}\right\}|\left\{{\bf r}_{2,\beta}\right\}) \,, \nonumber\\
\end{align} 
and
\begin{align}
j^x_{2,\beta}\left(\left\{{\bf r}_{1,\alpha}\right\}|\left\{{\bf r}_{2,\beta}\right\}\right) &= \Big[\left(B_N T_2 - E_N T_2 - \kappa - N u\right) x_{2,\beta}  + N \left(u  - \lambda_N C_N T_2\right) X_1+ \nonumber\\ &+ N E_N T_2 X_2 
\Big] P(\left\{{\bf r}_{1,\alpha}\right\}|\left\{{\bf r}_{2,\beta}\right\})  \nonumber\\
j^y_{2,\beta}\left(\left\{{\bf r}_{1,\alpha}\right\}|\left\{{\bf r}_{2,\beta}\right\}\right) &= \Big[\left(B_N T_2 - E_N T_2 - \kappa -N u\right) y_{2,\beta} +N  \left(u  - \lambda_N C_N T_2\right) Y_1 + \nonumber\\ &+ N E_N T_2 Y_2 
\Big]  P(\left\{{\bf r}_{1,\alpha}\right\}|\left\{{\bf r}_{2,\beta}\right\})  \,,
\end{align}
where $X_1$, $Y_1$, $X_2$ and $Y_2$ define the coordinates of the centers of masses of the two ensembles of particles, eqs. \eqref{cm}.

Correspondingly, the sums (over $\alpha$ and $\beta$) of the divergences of the currents obey :
	\begin{align}
		& \sum_{\alpha} {\rm div}_{1,\alpha} \, {\bf j}_{1,\alpha} = \Big[ 2 N \left(A_N T_1 - \kappa -N u\right) \nonumber\\
		&+ \left(D_N - A_N\right) \left(A_N T_1 - D_N T_1 -  \kappa - N u\right) \Big(\sum_{\alpha} x_{1,\alpha}^2 + \sum_{\alpha} y_{1,\alpha}^2\Big) \nonumber\\
		&+ N^2 \Big[ \lambda_N C_N \left(A_N T_1 - D_N T_1 - \kappa - N u\right) + \left(D_N - A_N\right) (u - \lambda_N C_N T_1) \nonumber\\&+ \lambda_N N C_N D_N T_1  - N D_N (u - \lambda_N C_N T_1)
		\Big] \left(X_1 X_2 + Y_1 Y_2\right) \nonumber\\
		&+ N^2 \Big[ (D_N - A_N) D_N T_1 - D_N (A_N T_1 - D_N T_1 - \kappa - N u) - N D_N^2 T_1
		\Big] \left(X_1^2 + Y_1^2\right) \nonumber\\
		& + \lambda_N N^3 C_N (u - \lambda_N C_N T_1) \left(X_2^2 + Y_2^2\right) \Big] P(\left\{{\bf r}_{1,\alpha}\right\}|\left\{{\bf r}_{2,\beta}\right\}) \,,
	\end{align}
and
\begin{align}
&\sum_{\beta} {\rm div}_{2,\beta} \, {\bf j}_{2,\beta} = \Big[ 2 N \left(B_N T_2 - \kappa - u\right) \nonumber\\
&+ \left(E_N - B_N\right) \left(B_N T_2 - E_N T_2 -  \kappa -N u\right) \Big(\sum_{\beta} x_{2,\beta}^2 + \sum_{\beta} y_{2,\beta}^2\Big) \nonumber\\
&+ N^2 \Big[ \lambda_N C_N \left(B_N T_2 - E_N T_2 - \kappa - N u\right) + \left(E_N - B_N\right) (u - \lambda_N C_N T_2) \nonumber\\&+ \lambda_N N C_N E_N T_2  - N E_N (u - \lambda_N C_N T_2)
\Big] \left(X_1 X_2 + Y_1 Y_2\right) \nonumber\\
&+ N^2 \Big[ (E_N - B_N) E_N T_2 - E_N (B_N T_2 - E_N T_2 - \kappa - N u) - N E_N^2 T_2
\Big] \left(X_2^2 + Y_2^2\right) \nonumber\\
& + \lambda_N N^3 C_N (u - \lambda_N C_N T_2) \left(X_1^2 + Y_1^2\right) \Big] P(\left\{{\bf r}_{1,\alpha}\right\}|\left\{{\bf r}_{2,\beta}\right\})  \,.
\end{align}
Eventually, in virtue of eq. \eqref{FPmany}, we have
\begin{align}
\sum_{\alpha} {\rm div}_{1,\alpha} \, {\bf j}_{1,\alpha} + \sum_{\beta} {\rm div}_{2,\beta} \, {\bf j}_{2,\beta} = 0,
\end{align}
which implies that

\begin{align}
& 0 = 2 N \Big[A_N T_1  + B_N T_2 - 2 (\kappa + N u)\Big] \nonumber\\
& + \Big[\left(D_N - A_N\right) \left(A_N T_1 - D_N T_1 - \kappa - N u\right) \Big] \left(\sum_{\alpha} x_{1,\alpha}^2 + \sum_{\alpha} y_{1,\alpha}^2\right) \nonumber\\& + \Big[\left(E_N - B_N\right) \left(B_N T_2 - E_N T_2-  \kappa - N u\right)\Big] \left(\sum_{\beta} x_{2,\beta}^2 + \sum_{\beta} y_{2,\beta}^2\right) \nonumber\\
&+N^2 \Big[\lambda_N C_N \left(A_N T_1 - D_N T_1 - \kappa - N u\right) + \left(D_N - A_N\right) \left(u - \lambda_N C_N T_1\right) \nonumber\\
&+ \lambda_N C_N \left(B_N T_2 - E_N T_2 - \kappa - N u\right) + \left(E_N - B_N\right) \left(u - \lambda_N C_N T_2\right) \nonumber\\
&+ 2 \lambda_N N C_N D_N T_1 - u N D_N  + 2 \lambda_N N C_N E_N T_2 - u N E_N
\Big] \left(X_1 X_2 + Y_1 Y_2\right) \nonumber\\
&+ N^2 \Big[ \left(D_N - A_N\right) D_N T_1 - D_N \left(A_N T_1 - D_N T_1 - \kappa -N u \right) \nonumber\\
&- N D_N^2 T_1 + \lambda_N N C_N \left(u - \lambda_N C_N T_2\right)
\Big] \left(X_1^2+Y_1^2\right) \nonumber\\
&+N^2 \Big[\left(E_N - B_N\right) E_N T_2 - E_N \left(B_N T_2 - E_N T_2 - \kappa - N u\right) \nonumber\\
&- N E_N^2 T_2 + \lambda_N N C_N \left(u - \lambda_N  C_N T_1\right) \Big] \left(X_2^2+Y_2^2\right)  \,.
\label{eq:Coeff}
\end{align}
Equating to zero the terms in square brackets appearing in front of functions of $X$ and $Y$, we obtain six equations for five unknown coefficients. Inspecting further these equation, we realize 
that 
\begin{align}
A_N T_1  + B_N T_2 &= 2 (\kappa + N u) \,, \nonumber\\
A_N T_1 - D_N T_1 &= \kappa + N u \,, \nonumber\\
B_N T_2 - E_N T_2 &= \kappa +  N u \,,
\end{align}
implying that $E_N T_2 = - D_N T_1$, which makes two last equations not independent. 
%Multiplying the two last equations of \ref{eq:Coeff } by $T_1$ and $T_2$, respectively, and taking  the difference between %two equations,  one obtains a linear equation.
%\begin{equation}
%\lambda N C_Nu (T_1-T_2)=2(\kappa+Nu)	D_NT_1
%\end{equation}
Solving then the remaining five equations we obtain the following explicit expressions for the coefficients

\begin{align}
A_N &= \frac{(\kappa+N u)}{T_1} +  \frac{(\kappa + N u)\lambda_N^2 N \left(T_1^2 - T_2^2\right)}{ T_1 \left(4 T_1 T_2 + \lambda_N^2 N^2 \left(T_1 - T_2\right)^2\right)}  \,, \nonumber\\
B_N &= \frac{(\kappa + N u)}{T_2} +  \frac{(\kappa + N u) \lambda_N^2 N \left(T_2^2 - T_1^2\right)}{T_2 \left(4 T_1 T_2 + \lambda_N^2 N^2 \left(T_1 - T_2\right)^2\right)}   \,, \nonumber\\
C_N  &= \frac{2 (\kappa+N u) \left(T_1 + T_2\right)}{ \left(4 T_1 T_2 + \lambda_N^2 N^2 \left(T_1 - T_2\right)^2\right)} \,, \nonumber\\
D_N & = \frac{(\kappa + N u) \lambda_N^2 N \left(T_1^2 - T_2^2\right)}{T_1 \left(4 T_1 T_2 + \lambda_N^2 N^2 \left(T_1 - T_2\right)^2\right)} \,, \nonumber\\
E_N &=    \frac{(\kappa + N u) \lambda_N^2 N \left(T_2^2 - T_1^2\right)}{T_2 \left(4 T_1 T_2 + \lambda_N^2 N^2 \left(T_1 - T_2\right)^2\right)}     \,,\label{eq:coeffs}
\end{align}  
which are valid for arbitrary $N$, $T_1$, $T_2$, $\kappa$ and $u$. Finally, performing the $4 N$-fold integrals, we find that the normalization is explicitly given by
%Note that the coefficients $A_N$ and $B_N$ are positive definite for any values of the parameters, which ensures the convergence of the integrals. 
%Setting $N = 1$, we recover from
% the latter expressions the explicit forms in eqs. \eqref{coefts}.
\begin{equation}
Z_N=\frac{(2\pi)^{2N}}{\Big[(B_N-E_N)^{N-1}(A_N-D_N)^{N-1}\left((A_N+(N-1)D_N)(B_N+(N-1)E_N)-N^2\lambda_N^2 C_N^2\right)\Big]}\label{eq:norm2}
\end{equation} 
Inserting eqs.\eqref{eq:coeffs} into eq.\eqref{eq:norm2} yields 
\begin{equation}
Z_N=\frac{{2}^{2N-2}\pi^{2N} (T_1T_2)^{N-1} \left(4 T_1 T_2 + \lambda_N^2 N^2 \left(T_1 - T_2\right)^2\right)}{(\kappa+Nu)^{2N}(1-N^2\lambda_N^2)} 
\end{equation}
such that one recovers eq.\eqref{eq:Z1}

In virtue of eq.\eqref{eq:modt},   eq.\eqref{eq:synchro} can be expressed as
\begin{equation}
|s|=\frac{2N\left(\kappa + N u\right) (1 - \lambda^2_N N ^2)  |u (T_1 - T_2)| }{4 T_1 T_2 + \lambda^2_N N^2 \left(T_1 - T_2\right)^2}\int_\infty^\infty d\omega \frac{1}{\pi\omega^2}\left(1-\frac{Z_N(\omega)}{Z_N}\right)
\end{equation}
where $Z_N(\omega)$ is the normalization of the integral of the function
\begin{equation}
P\left(\left\{{\bf r}_{1,\alpha}\right\}|\left\{{\bf r}_{2,\beta}\right\}\right) \exp(i\omega (X_1Y_2-X_2Y_1))
\end{equation}

\begin{align}
	Z_N(\omega)&=\frac{(2\pi)^{2N}}{\left[(B_N-E_N)^{N-1}(A_N-D_N)^{N-1}\left( (A_N+(N-1)D_N)(B_N+(N-1)E_N)
		-N^2\lambda_N^2 C_N^2+\frac{\omega^2}{N^2}\right)\right]}\label{eq:norm3}
\end{align} 
One then obtains that 
\begin{equation}
\frac{Z_N(\omega)}{Z_N}=\frac{1}{1+G^2_N\omega^2}
\end{equation}
where 
\begin{align}
G_N&=\frac{1}{N\sqrt{((A_N+(N-1)D_N)(B_N+(N-1)E_N)- N^2\lambda^2_NC^2_N)}}
\nonumber\\
&=\frac{1}{N}\sqrt{\frac{ \left(4 T_1 T_2 + \lambda_N^2 N^2 \left(T_1 - T_2\right)^2\right)}{4(1-N^2\lambda_N^2)(\kappa+Nu)^2}}\label{eq:GN}
\end{align}
Integrating over $\omega$ gives
\begin{align}
|s|=&\frac{2 N\left(\kappa + N u\right) (1 - \lambda^2_N N ^2)  |u (T_1 - T_2)| }{4 T_1 T_2 + \lambda^2_N N^2 \left(T_1 - T_2\right)^2}G_N\label{eq:sn}
\end{align}
Inserting eq.\eqref{eq:GN} into eq.\eqref{eq:sn} leads to the expression \eqref{eq:synchro}.

\section{Four-particle case}

In order to compare our analytical predictions and the results of numerical simulations, we focus on a particular case with $N=2$, meaning that the system comprises two particles of each of the two species.
%The system is now composed of 4 particles with two species.
%By using the general results, one can obtain the differents moments as well the absolute values of mean curl wich measures the intensity of the synchronisation.
For such a system, we first calculate 
the mean-squared inter-particle separation distance between dissimilar particles 
\begin{align}
	\overline{({\bf r}_{1,\alpha} - {\bf r}_{2,\beta})^2} &= \int \int d\{{\bf r}_{1,\alpha}\} d\{{\bf r}_{2,\beta}\} ({\bf r}_{1,\alpha} - {\bf r}_{2,\beta})^2 	P(\left\{{\bf r}_{1,\alpha}\right\}|\left\{{\bf r}_{2,\beta}\right\}) \nonumber\\&=  
	\frac{2 (T_1 + T_2)(\kappa+3u)}{(\kappa + 2 u)(\kappa+4u)} \,, \label{eq:rel4dis}
\end{align}
which signifies that the dissimilar particles become delocalised when $u \to -\kappa/4$. In turn, for the mean-squared separation distance between like particles we get 
\begin{align}
	\overline{({\bf r}_{i,\gamma} - {\bf r}_{i,\gamma'})^2} &= \int \int d\{{\bf r}_{1,\alpha}\} d\{{\bf r}_{2,\beta}\}  ({\bf r}_{i,\gamma} - {\bf r}_{i,\gamma'} )^2 P\left(\left\{{\bf r}_{1,\alpha}\right\}|\left\{{\bf r}_{2,\beta}\right\}\right) \nonumber\\&
	= \frac{4 T_i}{\kappa + 2 u} \,,  \label{eq:rel4dis2}
\end{align}
which implies that like particles get decoupled when $u \to -\kappa/2$.

\begin{figure}
	\begin{center}
		\includegraphics[width=70mm]{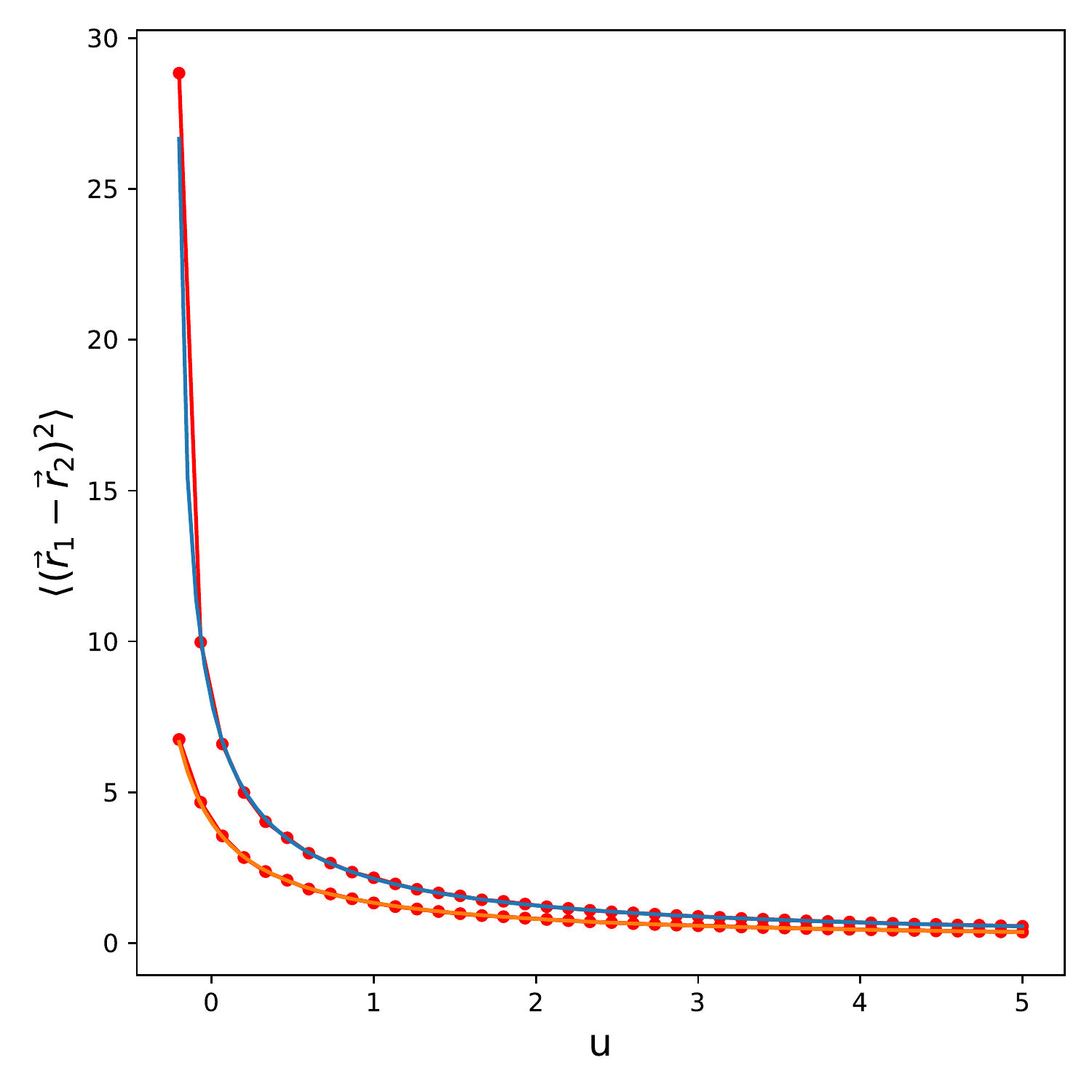}
		\includegraphics[width=70mm]{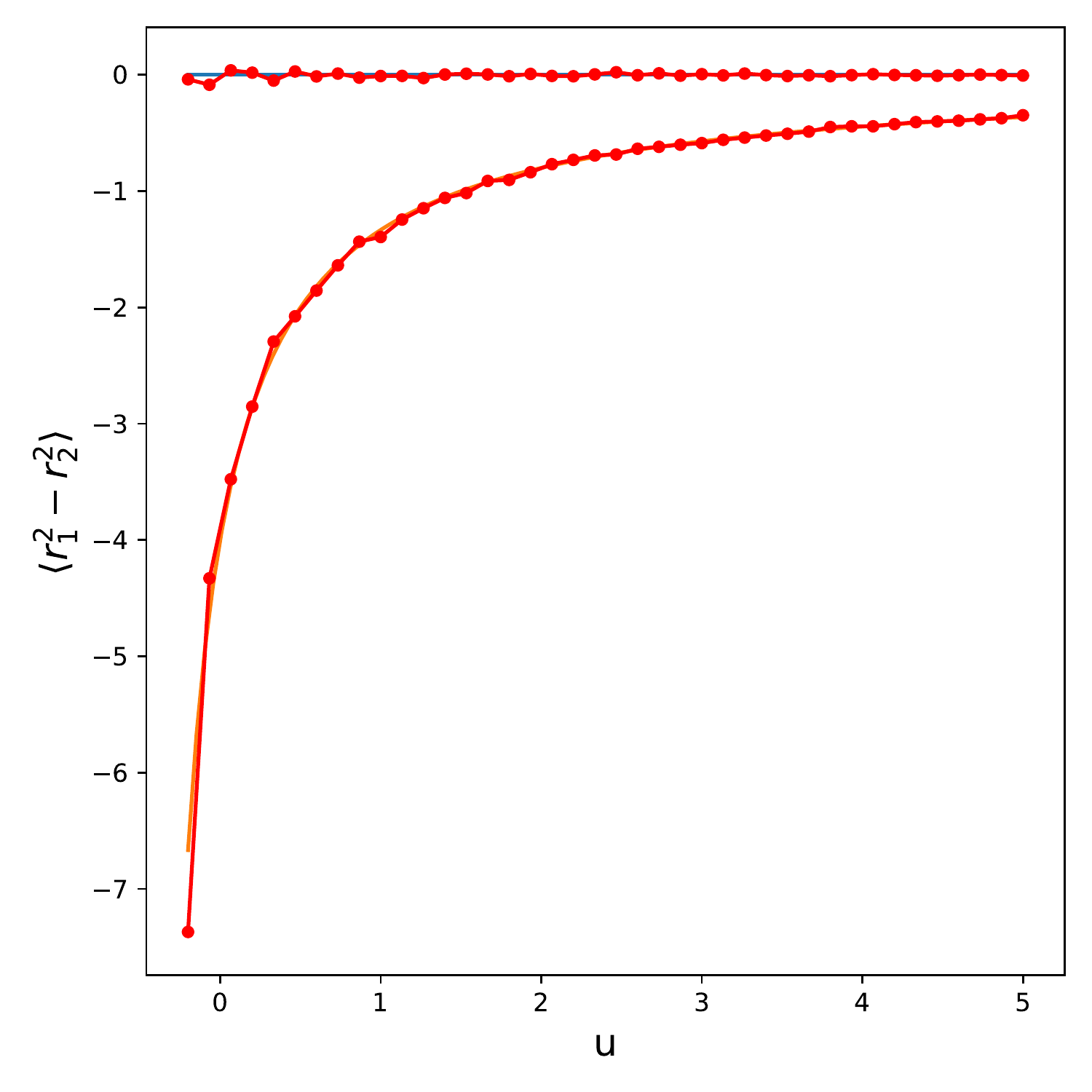}
	\end{center}
	\caption{4 particles-case: (left) Mean-squared inter-particle separation distance 
		and (right)  distance between the two particles  of different species (upper curve) and two particles of species $1$ (lower curve)  versus the strength of the interaction between particles $u$ for $T_1 = 1$ and
		$T_2 = 3$. 
		Full curves correspond to eq.~\eqref{eq:rel4dis}  and eq.~\eqref{eq:rel4dis2} (left) and eq.~\eqref{eq:rel4dis3} (right). Dots correspond to the  simulation results.}
	\label{fig:top4r}
\end{figure} 
Similarly, one finds that  the difference of the mean-squared displacements of the two particles of different species obeys
\begin{align}
	\overline{{\bf r}_{1,\alpha}^2 - {\bf r}_{2,\beta}^2} &= \int \int d\{{\bf r}_{1,\alpha}\} d\{{\bf r}_{2,\beta}\}( {\bf r}_{1,\alpha}^2 - {\bf r}_{2,\beta}^2) P\left(\left\{{\bf r}_{1,\alpha}\right\}|\left\{{\bf r}_{2,\beta}\right\}\right)  
	= \frac{2 (T_1 - T_2)}{\kappa + 2 u} \,. \label{eq:rel4dis3}
\end{align}
where as the difference of the mean-squared displacements of two particles of the same
specied vanishes.
\begin{align}
	\overline{{\bf r}_{i,\gamma} ^2 - {\bf r}_{i,\gamma'}^2} &= \int \int d\{{\bf r}_{1,\alpha}\} d\{{\bf r}_{2,\beta}\} ({\bf r}_{i,\gamma}^2 - {\bf r}_{i,\gamma'} ^2) P({\bf r}_1|{\bf r}_2) 
	= 0 \,. \label{eq:reldis44}
\end{align}
For measuring the intensity of the synchronisation, one can use Eq.\ref{eq:synchro}  for $N=2$ and the exact expression  is given by
\begin{equation}
	|s_{12}|=|u(T_1-T_2)|\sqrt{\frac{1-4 \lambda_2^2}{4T_1T_2+4\lambda^2_2(T_1-T_2)^2}}\label{eq:synchro4}
\end{equation}

\begin{figure}
	\begin{center}
		\includegraphics[width=90mm]{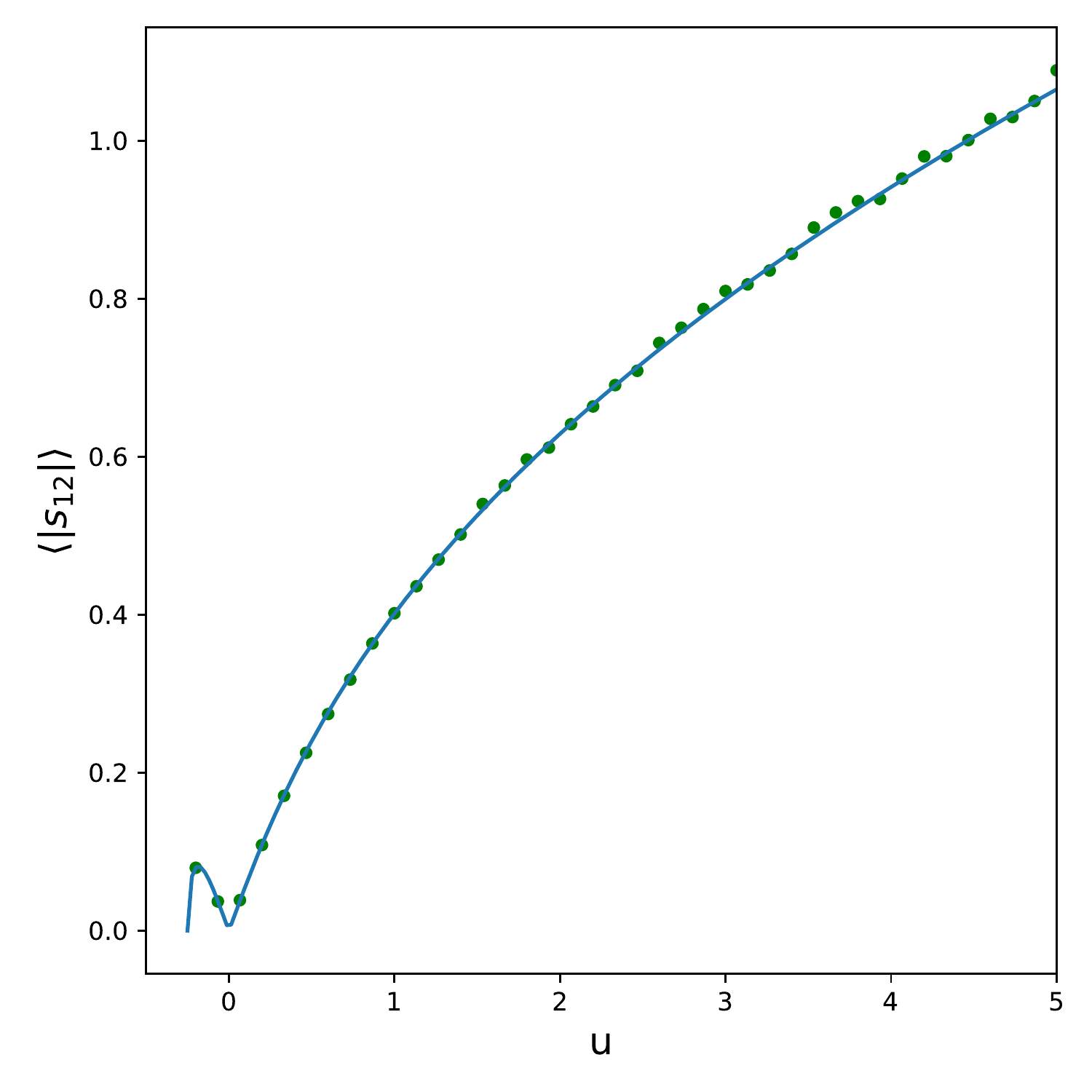}
	\end{center}
	\caption{$|s_{1,2}|$ as a function of $u$ for $T_1=1$ and $T_2=3$. Full curves correspond to  the exact expression (see eq.\eqref{eq:synchro4}) dots correspond to the simulation results for $2N=4$.
	}
	\label{fig:s412}
\end{figure} 
Figure \ref{fig:s412} shows the evolution of the absolute value of $s_{12}$ versus $u$. Compared to the $2$ particle case, the amplitude decreased.
It is worth noting that with $4$ particles the stability of the system is bounded for repulsive coupling at $u=-\kappa/4$ conversely to the 2 particle case where $u$ was equal to $-\kappa/2$.  Full curve corresponds to the analytical expression  Simulation results are displayed for $2N=4$ and match the analytical expression.

%\bibliography{synchro}

\end{document}